\documentclass[journal,twoside,web]{ieeecolor}
\usepackage{generic}
\usepackage{cite}
\usepackage{amsmath,amssymb,amsfonts}
\usepackage{algorithmic}
\usepackage{graphicx}
\usepackage{algorithm,algorithmic}
\usepackage{hyperref}
\hypersetup{hidelinks=true}
\usepackage{textcomp}

\usepackage{bm}

\def\BibTeX{{\rm B\kern-.05em{\sc i\kern-.025em b}\kern-.08em
    T\kern-.1667em\lower.7ex\hbox{E}\kern-.125emX}}
\begin{document}
 \title{Robust Model Predictive Control for Constrained Uncertain Systems Based on Concentric Container and Varying Tube}
	
\author{ 
	Shibo Han, Yuhao Zhang, Xiaotong Shi, Xingwei Zhao
\thanks{This work was supported in part by the China Scholarship Council under Grant 202106160015, and the National Natural Science Foundation of China under Grant 52422501. (Corresponding author: Shibo Han)}
\thanks{Shibo Han is with the Department of Mechanical Engineering, National University of Singapore, 117576 Singapore (e-mail: e0846825@n.nus.edu).}
\thanks{Xingwei Zhao and Yuhao Zhang are with the State Key Laboratory of Intelligent Manufacturing Equipment and Technology, Huazhong University of Science and Technology, Wuhan 430074, China (e-mail: zhaoxingwei@hust.edu.cn, yuhao\_zhang@hust.edu.cn).}
\thanks{Xiaotong Shi is with the School of Artificial Intelligence, Hubei University, Wuhan 430062, China (email:xiaotongshi@hubu.edu.cn).}
}

\maketitle

\begin{abstract}
	This paper proposes a novel robust model predictive control (RMPC) method for the stabilization of constrained systems subject to additive disturbance (AD) and multiplicative disturbance (MD). 
	Concentric containers are introduced to facilitate the characterization of MD, and varying tubes are constructed to bound reachable states.
	By restricting states and the corresponding inputs in containers with free sizes and a fixed shape, 
	feasible MDs, which are the products of model uncertainty with states and inputs, are restricted into polytopes with free sizes.
	Then, tubes with different centers and shapes are constructed based on the nominal dynamics and the knowledge of AD and MD.
	The free sizes of containers allow for a more accurate characterization of MD, 
	while the fixed shape reduces online computational burden, making the proposed method less conservative and computationally efficient.
	Moreover, the shape of containers is optimized to further reduce conservativeness. 
	Compared to the RMPC method using homothetic tubes, the proposed method has a larger region of attraction while involving fewer decision variables and constraints in the online optimization problem.

\end{abstract}


\textbf{This work has been submitted to the IEEE for possible publication. Copyright may be transferred without notice, after which this version may no longer be accessible.}

\ 

\section{Introduction}
\label{sec:introduction}

\IEEEPARstart{M}{odel} predictive control (MPC) is a widely-used control technique due to its ability to control systems with operating constraints \cite{kohler2024analysis}.
MPC uses the dynamic model of a given plant to predict the future behavior of the system. 
At each time step, MPC solves a constrained optimization problem which ensures the predicted states and inputs satisfy operating constraints and minimizes a given performance index. 
When the dynamic model is uncertain, which is the most general case, robust model predictive control (RMPC) is developed to take the uncertainty into account and prevent possible violation of constraints or even instability \cite{saltik2018outlook}. 

Uncertainties caused by additive disturbance (AD) and multiplicative disturbance (MD) have been extensively studied and RMPC methods have been proposed for uncertain systems with AD, MD, or both. 
AD arises from measurement noise and process noise, which is independent of states and inputs, 
while MD arises from plant-model mismatching and is determined by the product of unknown modeling error with states and inputs \cite{drgovna2020all}. 
Meanwhile, linear parameter varying (LPV) systems and nonlinear systems can be modeled as linear invariant systems under MD, and RMPC can be developed accordingly, 
	see \cite{bujarbaruah2021simple}, \cite{morato2020model}, \cite{pereira2018linear}, \cite{rakovic2022homothetic}, etc.
As AD is bounded and independent of states and inputs, its effects can be estimated and eliminated easily. 
Well-developed RMPC algorithms for systems subject to AD can be found in \cite{chisci2001systems},\cite{mayne2005robust}, \cite{rakovic2012homothetic}, \cite{sieber2021system}, etc.
In contrast, it is difficult to characterize MD in a less complex and less conservative manner since MD is related to states and inputs that change over time.  
Thus, developing RMPC methods for systems subject to MD is challenging and receives sustained attention.
 
For uncertain systems with MD, a variety of RMPC methods have been proposed, including 
linear-matrix-inequality-based (LMI-based) methods, 
uncertainty-over-approximation-based (UOA-based) method, 
and tube-based methods. 
LMI-based methods employ an affine state feedback policy obtained by solving a convex optimization problem involving linear matrix inequalities, see \cite{kothare1996robust}, \cite{kouvaritakis2000efficient}, \cite{houska2016short}, \cite{georgiou2022computationally}. 
UOA-based methods align time-varying state feedback gain to current and predicted states, see \cite{bujarbaruah2021simple},  \cite{rakovic2022homothetic},  \cite{bujarbaruah2022robust}, \cite{chen2024robust}.
Tube-based control laws consist of a predefined state feedback and a free control variable to be optimized, see \cite{fleming2013regions}, \cite{lorenzen2019robust}, \cite{kohler2019linear}.
Among those three types of methods, tube-based methods have the lowest computational burden \cite{chen2024robust} and do not require much offline parameter design, 
making it the central concern of this paper.

One promising tube-based RMPC structure for systems with MD is proposed in \cite{fleming2013regions} and \cite{fleming2014robust}, 
which is inspired by the homothetic-tube-based (HT-based) RMPC method introduced in \cite{rakovic2012homothetic}. 
Homothetic tubes with free sizes and a fixed shape are employed to bound the system trajectories, introducing additional flexibility to handle the state- and input-dependent MD.
It is demonstrated that RMPC methods using homothetic tubes outperform methods using tubes with fixed shape and size.
This method is further developed in \cite{lorenzen2019robust}, in which MD and AD are considered simultaneously.
Based on \cite{lorenzen2019robust}, an RMPC-based explicit dual control algorithm is proposed in \cite{parsi2022explicit} to address the robust constant reference tracking problem for systems subject to AD and MD.
In these methods, the centers of tubes are decision variables, which results in a heavy computational burden when a large prediction horizon is used or a high-dimension system is considered.
To reduce computational burden, the centers of tubes in \cite{kohler2019linear} evolve according to the nominal dynamics. However, the region of attraction of \cite{kohler2019linear} shrinks, as well.
On the other hand, tubes with a fixed shape overestimate the set of reachable states and introduce conservativeness.
Then, tubes formed by half-planes with free offsets are employed in \cite{lu2019robust} and \cite{heydari2021robust}. 
However, these approaches introduce a number of additional decision variables and a larger region of attraction is achieved at the cost of greater computational complexity.

As the uncertainty in predicted states evolves over time, another class of RMPC methods utilizes varying tubes to bound the system trajectories, 
referred to as the varying-tube-based (VT-based) method. 
The centers and shapes of tubes evolve according to the nominal dynamics and the knowledge of admissible disturbance \cite{chisci2001systems}, \cite{mayne2009robust}.
It is demonstrated in \cite{zanon2021similarity} that the VT-based method has a larger region of attraction than the method using tubes with fixed shape and size.
With admissible MD determined, varying tubes can be constructed and VT-based RMPC methods can be developed to solve the robust stabilization problem of systems with both AD and MD.
One attempt is given in \cite{bumroongsri2017robust} where the admissible MD is significantly overestimated and restricted in a time-invariant set, making this method very conservative.
Based on the assumption that MD resides within a compact ball whose radius can be derived from the state with a known $\mathcal{K} $-function, a VT-based RMPC algorithm is proposed in \cite{pin2009robust}.
Similarly, the VT-based RMPC in \cite{kohler2020computationally} constructs tubes online under the assumption that the upper bound on the uncertainty near the predicted states and inputs is known.
However, these two methods rely on strong assumptions about MD. 
Meanwhile, the online construction of tubes, along with the consequent restriction of constraints on nominal systems, is computationally intensive and should be avoided.

In this paper, a novel RMPC method using concentric containers and varying tubes, donated as CC-VT RMPC, is developed to address the stabilization problem of constrained systems subject to both AD and MD. 
The key challenge in utilizing varying tubes is estimating admissible MD less conservatively while maintaining a low computational burden, which is addressed by introducing concentric containers.  
Since MD is the product of model uncertainty with predicted states and inputs, 
	where model uncertainty belongs to a known bounded set while states and inputs change over time, 
	admissible MD can be effectively estimated if states and inputs reside within specific sets.
Thus, the proposed method introduces additional constraints to restrict predicted states and inputs within containers with free sizes and a fixed shape.
Free sizes allow for a tighter characterization of MD, while the fixed shape reduces online computational burden.
As a result, MDs are restricted within bounded sets with varying sizes and a fixed shape.
Tubes are constructed accordingly and CC-VT RMPC is developed then.
The main contributions of this paper are summarized as follows.
\begin{itemize}
	\item A novel RMPC method, CC-VT RMPC, is proposed with the ability to deal with MD effectively, providing a novel solution to the stabilization problem of constrained systems subject to both AD and MD.
	\item Concentric containers are introduced to RMPC structure to approximate admissible MD in a computationally efficient and less conservative manner.
	\item Method to optimize the shape of containers is proposed, which enlarges the region of attraction of the proposed method and further reduces conservativeness.
	\item A recursive algorithm is proposed to determine terminal constraints which guarantees robust recursive feasibility of the proposed method under both AD and MD. 
\end{itemize}

The rest of this paper is organized as follows. 
Section II  formulates the problem and presents some properties of set operations. 
Section III introduces the VT-based structure, additional constraints, terminal constraints, and the theoretical properties of the proposed CC-VT RMPC algorithm.
Section IV  outlines the method to optimize containers.
Section V   provides examples to demonstrate the effectiveness of the proposed RMPC algorithm.
Section VI  compares the proposed CC-VT RMPC algorithm with the HT-based algorithm.
Section VII concludes this paper.

\textbf{Notation:}
	The sets of real numbers and integers are denoted as $\mathbb{R}$ and $\mathbb{N}$, respectively. 
	$\mathbb{N}_{[a,b]} = \{x \in \mathbb{N} | a \leq x \leq b \}$.
	The sets of positive and non-negative integers are denoted as $\mathbb{N}^+$ and $\mathbb{N}^\dagger$, respectively.
	The unit closed ball in $n$ dimensions centered in the origin is denoted as $\mathbb{B}^n$.
	$I$ denotes the identity matrix with appropriate dimensions.
	The $i$th element of a vector $x$ and the $i$th row of matrix $A$ are notated as $[x]_i$ and $[A]_i$, respectively.
	$\left \| x \right \| _\psi^2 = x'\psi x$.
	$\psi$ is positive definite if $\left \| x \right \| _\psi^2 >0, \forall x \neq 0$.
	A positive definite $\psi$ is denoted as $\psi \succ 0$.
	A polytopic set is a convex set which is expressed as $\mathbb{X} = \{x|H_{ieq}x \leq h_{ieq} \}$. 
	Polytope, which is a bounded polytopic set, can be represented by the convex hull of vertices. 
	The convex hull of vertices $x_1,x_2,...,x_n$ is denoted by $\mathbb{CH}\{x_1,x_2,...,x_n\}$. 
	For sets $\mathbb{A}\subseteq\mathbb{R}^{m\times n},\mathbb{X}\subseteq\mathbb{R}^n$,  matrix $A \in \mathbb{R}^{m\times n}$, and vector $x \in \mathbb{R}^n$, define $\mathbb{A}\mathbb{X}=\{Ax|A\in\mathbb{A},x\in\mathbb{X}\}$, $\mathbb{A}x=\{Ax|A\in\mathbb{A}\}$, and $A\mathbb{X}=\{Ax|x\in\mathbb{X}\}$.

\section{Problem formulation and preliminaries}
	\label{sec:problemFormulation}
Consider the discrete-time linear system described by 
\begin{align}  \label{eqn:actualSystem}
	x(t+1)=A_n x(t)+B_n u(t)+w(t)+w_{\mathrm{M}}(t),
\end{align}
where $x(t)\in\mathbb{R}^n,u(t)\in\mathbb{R}^m,w(t)\in\mathbb{R}^n,w_{\mathrm{M}}(t)$ are state, control input, additive disturbance (AD), and multiplicative disturbance (MD), respectively. 
$A_n$ and $B_n$ are matrices with compatible dimensions, representing the known nominal dynamics. MD is caused by unknown model mismatching, which is given as
\begin{align}  
	w_{\mathrm{M}}=\Delta P \begin{bmatrix}x(t)\\u(t)\end{bmatrix},
\end{align}
where $\Delta P = \begin{bmatrix}\Delta A & \Delta B\end{bmatrix}$ represents unknown modeling error. When (\ref{eqn:actualSystem}) is used to characterize a linear time invariant system with model mismatching, $\Delta P$ is constant. While when (\ref{eqn:actualSystem}) is used to characterize a LPV system or nonlinear system, $\Delta P$ is time-varying.

Meanwhile, there are constraints on state and input of system (\ref{eqn:actualSystem}), which is expressed as
\begin{align}  \label{eqn:systemConstraints}
	\begin{bmatrix}x(t)\\u(t)\end{bmatrix}\in\mathbb{Z} \subset \mathbb{R}^{n+m}.
\end{align}

Assumptions on constraints and uncertainties of system (\ref{eqn:actualSystem}) are given below.

(\textbf{A1}) $\mathbb{Z}$ is a polytopic set containing the origin in its interior.

(\textbf{A2}) $w(t) \in \mathbb{W}$ where $\mathbb{W} \subset \mathbb{R}^n$ is a polytope containing the origin in its interior.

(\textbf{A3}) $\Delta P \in \mathbb{P}_\Delta$ where $\mathbb{P}_\Delta \subset \mathbb{R}^{n \times (n+m)}$ is a polytope containing the origin in its interior.

(\textbf{A4}) There exists a feedback gain $K$ such that  $(A_n+\Delta A)+(B_n+\Delta B)K$ is Schur for all $\begin{bmatrix}\Delta A & \Delta B\end{bmatrix} \in \mathbb{P}_\Delta$.

With the given $K$, $A_n^{cl}$ is define as $A_n^{cl} \triangleq A_n+B_nK$.

As $\mathbb{P}_\Delta$ is a polytope, it can be expressed as 
\begin{align} 
	\mathbb{P}_\Delta = \mathbb{CH}\{ \Delta P_i ^{\mathrm{vert}} \},i\in\mathbb{N}_{\left[1,N_{\mathbb{P}_\Delta}^{\mathrm{vert}} \right]}.
\end{align}

$\Delta P_i ^{\mathrm{vert}}$ is the $i$th vertex of polytope $\mathbb{P}_\Delta$, and $N_{\mathbb{P}_\Delta}^{\mathrm{vert}}$ is the number of the vertices of polytope $\mathbb{P}_\Delta$. 
In the following parts, the vertices and the number of vertices of a polytope are notated in a similar way.

Define   
\begin{align}  
	\mathbb{W}_\mathrm{M}\big(x(t),u(t)\big) \triangleq \mathbb{P}_\Delta\begin{bmatrix}x(t)\\u(t)\end{bmatrix}.
\end{align}

 As state and input change over time, $\mathbb{W}_\mathrm{M}\big( x(t),u(t) \big)$ is time-varying. 
 Even worse, because of the existence of AD and MD, future states cannot be predicted accurately,
 bringing great challenges to the estimation of MD and the design of a RMPC controller.

The objective of this paper is to design a nonlinear feedback controller $\pi(x(t))$ which guarantees the robust stability as well as the robust constraint satisfaction of system (\ref{eqn:actualSystem}) for all admissible AD and MD satisfying (A2), (A3), and (A4).

Some preliminaries of set theory used in this paper are given below. 
For non-empty sets $\mathbb{X}_1 \subset \mathbb{R}^n, \mathbb{X}_2 \subset \mathbb{R}^n $, the Minkowski set addition and Pontryagin set difference are defined by 
$\mathbb{X}_1 \oplus \mathbb{X}_2 = \{ x_1 + x_2 | x_1 \in \mathbb{X}_1, x_2 \in \mathbb{X}_2 \}$ 
and $\mathbb{X}_1 \ominus \mathbb{X}_2 = \{ x| x \oplus \mathbb{X}_2 \subseteq \mathbb{X}_1 \}$, respectively.
If $\mathbb{X}_2$ contains only one element $x_2$, these expressions simplify to $\mathbb{X}_1 \oplus x_2$ and $\mathbb{X}_1 \ominus x_2$.

For non-empty convex sets $\mathbb{X}, \mathbb{X}_1$ and $\mathbb{X}_2$, 
\begin{align*}  
	\mathbb{X}_{1}\oplus \mathbb{X}_{2}   &=\mathbb{X}_{2}\oplus \mathbb{X}_{1}  \\
	\mathbb{X}\ominus \mathbb{X}_{1}\ominus \mathbb{X}_{2}	&=\mathbb{X}\ominus \mathbb{X}_{2} \ominus \mathbb{X}_{1}  \\
	T \mathbb{X}_1 \oplus T \mathbb{X}_2  &= T(\mathbb{X}_1 \oplus \mathbb{X}_2) \\
	T \mathbb{X}_1 \ominus T \mathbb{X}_2 &= T(\mathbb{X}_1 \ominus \mathbb{X}_2) \\
	\mathbb{X}_1\oplus \mathbb{X}_2\ominus \mathbb{X}_2 &=  \mathbb{X}_1  \\
	\mathbb{X}_1\ominus \mathbb{X}_2\oplus \mathbb{X}_2 &\subseteq \mathbb{X}_1 \\
	(\mathbb{X}_1 \cap \mathbb{X}_2) \ominus \mathbb{X} &= (\mathbb{X}_1 \ominus \mathbb{X}) \cap (\mathbb{X}_2 \ominus \mathbb{X}).
\end{align*}

Further, if $\mathbb{X}_1 \subseteq \mathbb{X}_2$, then
\begin{align*}  
	T \mathbb{X}_1    &\subseteq T \mathbb{X}_2 \\
	\mathbb{X}\oplus  \mathbb{X}_1 &\subseteq \mathbb{X}\oplus \mathbb{X}_2 \\
	\mathbb{X}\ominus \mathbb{X}_1 &\supseteq \mathbb{X}\ominus \mathbb{X}_2.	
\end{align*}

\section{Formulation of CC-VT RMPC}
In this section, the proposed CC-VT RMPC method is presented. 
With the analysis of the deviation between the nominal and actual systems, the VT-based structure is developed.
Then, additional constraints are introduced to restrict predicted states and inputs within containers. 
After that, terminal constraints for systems under both AD and MD are determined.
Finally, the proposed CC-VT RMPC method is concluded and its properties are proved theoretically.

\subsection{VT-based Structure}
The nominal system corresponding to system (\ref{eqn:actualSystem}) is 
\begin{align}  \label{eqn:nominalSystem}
	\bar{x}(t+1)=A_n \bar{x}(t)+B_n\bar{u}(t).
\end{align}

To restrict the deviation between the actual and nominal systems, the control inputs to those two systems are chosen as 
\begin{align}
	 u(t)=Kx(t)+v(t),  \label{eqn:actualControlInput}\\
	 \bar{u}(t)=K\bar{x}(t)+v(t), \label{eqn:nominalControlInput}
\end{align}
where $v(t)$ is a free control variable to be determined.

Suppose at $t$, the predicted control variables at $t+k$ is chosen as $v(k|t), k \in \mathbb{N}^\dagger$. 
With initial condition 
\begin{align}		\label{eqn:initialCondition}
	x(0|t)=x(t), \bar{x}(0|t)=x(t), \mathbb{E}(0|t) = \{0\},
\end{align}
and control law defined above, 
the predicted nominal states $\bar{x}(k|t)$ in the following steps can be determined according to (\ref{eqn:nominalSystem}), 
while the actual state $x(k|t)$ remains uncertain because of AD and MD. 
Moreover, the uncertainty in predicted states and inputs of actual systems makes the characterization of MD even more challenging.

Notate $e(k|t) = x(k|t)-\bar{x}(k|t)$, we have 
\begin{align}  
	\label{eqn:accurate_e}
	e(k+1|t)=A_n^{cl} e(k|t)+w(k|t)+w_\mathrm{M}(k|t),
\end{align}
where $w_\mathrm{M}(k|t) = \Delta P \begin{bmatrix}x(k|t)\\u(k|t)\end{bmatrix} \in \mathbb{P}_\Delta \begin{bmatrix}x(k|t)\\u(k|t)\end{bmatrix}$. 

Suppose  $e(k|t)\in\mathbb{E}(k|t)$, since $u(k|t)=Kx(k|t)+v(k|t)=\bar{u}(k|t)+Ke(k|t)$, we have 
\begin{align}  \label{eqn:xuE}
	\begin{bmatrix}x(k|t)\\u(k|t)\end{bmatrix}\in\begin{bmatrix}\bar{x}(k|t)\\\bar{u}(k|t)\end{bmatrix}\oplus\begin{bmatrix}I\\K\end{bmatrix}\mathbb{E}(k|t).
\end{align}
Then, $w_\mathrm{M}(k|t) \in \mathbb{W}_\mathrm{M}(k|t)$ where
\begin{align}		\label{eqn:w_p_i_t}
	\mathbb{W}_\mathrm{M}(k|t) 
	= \mathbb{P}_\Delta \left( \begin{bmatrix}\bar{x}(k|t)\\\bar{u}(k|t)\end{bmatrix} \oplus         
	\begin{bmatrix}I\\K\end{bmatrix}\mathbb{E}(k|t)\right   ).
\end{align}

Equation (\ref{eqn:accurate_e}) points out how the deviation between the actual system and nominal system evolves, that is, 
\begin{align}  
	\label{eqn:accurateE}
	\mathbb{E}(k+1|t)=A_n^{cl}\mathbb{E}(k|t)\oplus\mathbb{W}\oplus \mathbb{W}_\mathrm{M}(k|t).
\end{align}

Although (\ref{eqn:w_p_i_t}) is computationally inefficient, it provides an optional bounded set which contains $w_\mathrm{M}(k|t)$,
making it possible to determine  $\mathbb{E}(k|t)$ recursively along with (\ref{eqn:accurateE}).

$\bar{x}(k|t) \oplus \mathbb{E}(k|t)$, the so-called "tube", is the set of states of the uncertain system (\ref{eqn:actualSystem}) which is reachable in $k$ steps. 
Tubes evolve over time and vary in centers and shapes.  
With (\ref{eqn:nominalSystem}) and (\ref{eqn:accurateE}), the evolution of tubes is pointed out.
A RMPC-based control law is then determined as 
\begin{align}  \label{eqn:controlLaw}
	\pi(x(t))=Kx(t)+v^*(0|t),
\end{align}
where $v^*(0|t)$ is obtained from the following finite horizon optimization problem $\mathbb{FHOP}(x(t))$.
\begin{align}
	\label{eqn:trivialMPC}
	\min_{\bm{v}(t)}   \quad  J(\bm{v}(t))
\end{align}
subject to
\begin{subequations} \label{eqn:QP_constraints_2001}
	\begin{align}
		\bar{x}(0|t)   &= x(t)       \\            
		\bar{x}(k+1|t) &= A_n\bar{x}(k|t) + B_n \bar{u}(k|t)         \\  
		\begin{bmatrix}\bar{x}(k|t)\\\bar{u}(k|t)\end{bmatrix} &\in 
					\mathbb{Z}\ominus\begin{bmatrix}I\\K\end{bmatrix}\mathbb{E}(k|t)   \label{eqn:feasibleConstraint_2011}\\
		\bar{x}(N|t)  &\in \mathbb{S}_\infty\ominus\mathbb{E}(N|t)  \\
		k             &\in \mathbb{N}_{[0,N-1]}         					
	\end{align}	
\end{subequations}
with 
\begin{subequations}
\begin{align}  
	\mathbb{E}(0|t)   &= \{0\} \\
	\mathbb{E}(k+1|t) &= A_n^{cl}\mathbb{E}(k|t)\oplus\mathbb{W}\oplus\mathbb{W}_\mathrm{M}(k|t)  \label{eqn:E_E_W_WP_2001}  \\
	\mathbb{W}_\mathrm{M}(k|t) &= \mathbb{P}_\Delta \left(\begin{bmatrix}\bar{x}(k|t)\\\bar{u}(k|t)\end{bmatrix} \oplus 
						 \begin{bmatrix}I\\K\end{bmatrix}\mathbb{E}(k|t)\right) \\ 		\label{eqn:WP_kt_2001}
	\bar{u}(k|t)      &= K \bar{x}(k|t) + v(k|t).
\end{align}
\end{subequations}

$\bm{v}(t)=[v'(0|t),v'(1|t),...,v'(N-1|t)]'$.
$N$ is the prediction horizon. 
Cost function $J(\bm{v}(t))$ and terminal constraints $\mathbb{S}_\infty$ are determined in the subsequent sections.

\subsection{Additional Constraints}
	\label{sec:restricedConstraints}
	To ensure robust constraint satisfaction, admissibility constraints (\ref{eqn:systemConstraints}) should be restricted according to (\ref{eqn:feasibleConstraint_2011}).
	However, $\mathbb{E}(k|t)$ have to be calculated online and iteratively according to (\ref{eqn:E_E_W_WP_2001}), making $\mathbb{FHOP}(x(t))$ (\ref{eqn:trivialMPC}) computationally intractable.
	This is the key challenge in stabilizing constrained systems with both AD and MD using RMPC methods. 
	 
	 For a polytope $\mathbb{Z}_m$ which contains the origin in its interior and $\lambda \geq 0$, if $z\in \mathbb{Z}_m$, then $\lambda z \in \lambda \mathbb{Z}_m$.
	 By definition, we have $\mathbb{P}_\Delta (\lambda \mathbb{Z}_m) = \lambda  \mathbb{P}_\Delta \mathbb{Z}_m$.
	 Then, for all sets $\mathbb{Z}_\mathrm{tube} \subseteq \lambda \mathbb{Z}_m$,
	 $\mathbb{P}_\Delta \mathbb{Z}_\mathrm{tube} \subseteq \lambda  \mathbb{P}_\Delta \mathbb{Z}_m$,
	 where $\mathbb{P}_\Delta \mathbb{Z}_m$ can be determined offline. 
	 
	 This inspires the introduction of additional constraints to make it computationally efficient to determine $\mathbb{W}_\mathrm{M}(k|t)$. 
	 With the condition that 
	 \begin{align} 
	 	\label{eqn:xBar_Zm_E}
	 	\begin{bmatrix}\bar{x}(k|t)\\K\bar{x}(k|t)+{v}(k|t)\end{bmatrix} \oplus \begin{bmatrix}I\\K\end{bmatrix}\mathbb{E}(k|t)
	 	 \subseteq\lambda(k|t)\mathbb{Z}_m,
	 \end{align}
	 we have $\mathbb{W}_\mathrm{M}(k|t) \subseteq \lambda(k|t) \mathbb{P}_\Delta \mathbb{Z}_m$. 
	 
	 $\lambda(k|t) \mathbb{Z}_m$ are termed "containers" since predicted states and control inputs of the actual system are contained in them. 
	 Containers are concentric. 
	 The shape of containers, $\mathbb{Z}_m$, is referred to as container, as well, for the sake of conciseness.
	 
	 Notate $\overline{\mathbb{W}}_\mathrm{M}(k|t)  = \lambda(k|t) \mathbb{P}_\Delta \mathbb{Z}_m$. 
	 Given $\overline{\mathbb{E}}(k|t)$ such that $\mathbb{E}(k|t)\subseteq\overline{\mathbb{E}}(k|t)$, 
	 with condition (\ref{eqn:xBar_Zm_E}), we have
	 \begin{align}
	 	\begin{aligned}
	 	\mathbb{E}(k+1|t) &= A_n^{cl}\mathbb{E}(k|t)\oplus\mathbb{W}\oplus\mathbb{W}_\mathrm{M}(k|t) \\
	 					  &\subseteq A_n^{cl} \overline{\mathbb{E}}(k|t) \oplus \mathbb{W}\oplus \overline{\mathbb{W}}_\mathrm{M}(k|t).
	 	\end{aligned}
	 \end{align} 
	 
Then, if condition (\ref{eqn:xBar_Zm_E}) holds for $k\in \mathbb{N}_{[0,N-1]}, \mathbb{E}(0|t)=\overline{\mathbb{E}}(0|t)=\{0\}$, 
and $\overline{\mathbb{E}}(k+1|t) = A_n^{cl}\overline{\mathbb{E}}(k|t)\oplus\mathbb{W}\oplus\overline{\mathbb{W}}_\mathrm{M}(k|t)$,
it can be concluded that $\mathbb{E}(k|t)\subseteq\overline{\mathbb{E}}(k|t),k\in \mathbb{N}_{[0,N]}$. 
Then, the constraints (\ref{eqn:QP_constraints_2001}) is further restricted to 
\begin{subequations} \label{eqn:QP_constraints}
	\begin{align}
		\bar{x}(0|t)   &= x(t)          \label{eqn:initialConstraints}            \\
		\bar{x}(k+1|t) &= A_n\bar{x}(k|t) + B_n \bar{u}(k|t)       \label{eqn:dynamicConstraint}\\
		\begin{bmatrix}\bar{x}(k|t)\\\bar{u}(k|t)\end{bmatrix} &\in 	
				 	\lambda(k|t) \mathbb{Z}_m \ominus \begin{bmatrix}I\\K\end{bmatrix}\overline{\mathbb{E}}(k|t)          	\label{eqn:addFeasibleConstraint}\\
		\begin{bmatrix}\bar{x}(k|t)\\ \bar{u}(k|t)\end{bmatrix} &\in 
			\mathbb{Z}\ominus\begin{bmatrix}I\\K\end{bmatrix}\overline{\mathbb{E}}(k|t)		\label{eqn:feasibleConstraint}\\
		\bar{x}(N|t)  &\in \mathbb{S}_\infty \ominus \overline{\mathbb{E}}(N|t)  \label{eqn:terminalConstraint}\\
		\lambda(k|t)  & \geq 0 		     \label{eqn:positivenessConstraint}		\\       
		k &\in \mathbb{N}_{[0,N-1]}      \label{eqn:count}   					
	\end{align}	
\end{subequations}
with 
\begin{subequations} \label{eqn:QP_constraints_equality}
	\begin{align}  
		\overline{\mathbb{E}}(0|t)   &= \{0\} 	\label{eqn:E_0}\\
		\overline{\mathbb{E}}(k+1|t) &= A_n^{cl}\overline{\mathbb{E}}(k|t) \oplus \mathbb{W}\oplus\overline{\mathbb{W}}_\mathrm{M}(k|t)  \label{eqn:E_E_W_WP}  \\
		\overline{\mathbb{W}}_\mathrm{M}(k|t)  &= \lambda(k|t) \mathbb{P}_\Delta \mathbb{Z}_m 		\label{eqn:WP_kt}  \\
		\bar{u}(k|t)      &= K \bar{x}(k|t) + v(k|t).  \label{eqn:uBar}		
	\end{align}
\end{subequations}

\textbf{Remark 1:}
Even with a number of free variables, it is difficult to determine a uniform parameterized linear-constraint expression or vertex expression of $\mathbb{W}_\mathrm{M}(k|t)$. 
Thus, to ensure the robust stability and the robust constraint satisfaction of system (\ref{eqn:actualSystem}) for all admissible MD, 
it is unavoidable to overestimate $\mathbb{W}_\mathrm{M}(k|t)$ as a compromise between computational complexity and conservativeness.

\textbf{Remark 2:}
In the proposed method, MD is restricted in $\lambda(k|t) \mathbb{P}_\Delta \mathbb{Z}_m$, which is a superset of $\mathbb{W}_\mathrm{M}(k|t)$.
It should be pointed out that conservativeness arises from $\lambda(k|t) \mathbb{P}_\Delta \mathbb{Z}_m$ rather than additional constraints (\ref{eqn:addFeasibleConstraint}). The feasible region of $\mathbb{FHOP}(x(t))$ (\ref{eqn:trivialMPC}) remains the same with constraints (\ref{eqn:addFeasibleConstraint}) been added.
Here, with decision variables $\lambda(k|t)$, additional constraints (\ref{eqn:addFeasibleConstraint}) provide information about the predicted states and inputs, 
	rather than imposing extra hard constraints on the system (\ref{eqn:actualSystem}) which restrict feasible region.

\subsection{Terminal Constraints}
Terminal constraints play a significant role in ensuring recursive feasibility. 
Terminal constraints $\mathbb{S}_\infty$ are supposed to satisfy the following assumption. 

(\textbf{A5}) For all $x\in \mathbb{S}_\infty$,  

(i) $x \in \mathbb{X}_{xu}$, where $\mathbb{X}_{xu} =\{x|\begin{bmatrix} x \\ Kx \end{bmatrix}  \in \mathbb{Z}\}$;

(ii) $\mathbb{S}_\infty \subseteq \gamma_\infty \mathbb{X}_m$, 
		where $\mathbb{X}_{m}=\Big\{x|\begin{bmatrix} x \\ Kx \end{bmatrix} \in \mathbb{Z}_m\Big\}$, 
		      $\gamma_\infty = \min_\gamma \{\gamma | \mathbb{S}_\infty\subseteq\gamma\mathbb{X}_m \}$;
		      
(iii) $A_n^{cl} x \oplus \mathbb{W} \oplus \gamma_\infty \mathbb{P}_\Delta \mathbb{Z}_m \subseteq \mathbb{S}_\infty$.

Requirement (i) ensures the admissibility of $\begin{bmatrix} x' & (Kx)' \end{bmatrix}'$.
Requirement (ii) guarantees that when $x \in  \mathbb{S}_\infty$ and $u=Kx$, 
admissible MD, that is, $\Delta P \begin{bmatrix} x' & (Kx)' \end{bmatrix}'$, resides within  $\gamma_\infty \mathbb{P}_\Delta \mathbb{Z}_m$, which is consistent with the invariant property of $\mathbb{S}_\infty$ given in requirement (iii).
$\mathbb{S}_\infty$ is the so-called "output admissible set" in \cite{kolmanovsky1998theory}.

Actually, $\gamma_\infty$ is unavailable before $\mathbb{S}_\infty$ is determined. 
Thus, we have to begin from an estimate of $\gamma_\infty$, notated as $\gamma_0$, to obtain $\mathbb{S}_n$.	
Then, the $0$-step admissible set is given as $\mathbb{S}_0=\gamma_0\mathbb{X}_m\cap\mathbb{X}_{xu}$. 
The $n$-step admissible set is notated as $\mathbb{S}_n, n\in \mathbb{N}^+$, which is determined recursively by
\begin{align} \label{eqn:Srecursive}
	\mathbb{S}_n = \{ x | x \in \mathbb{S}_0,  A_n^{cl} x \oplus \mathbb{W} \oplus \gamma_0 \mathbb{P}_\Delta \mathbb{Z}_m \subseteq \mathbb{S}_{n-1}\}.  
\end{align}

\textbf{Property 1:}
$\mathbb{S}_\infty \subseteq \mathbb{S}_{n+1} \subseteq \mathbb{S}_{n} \subseteq \mathbb{S}_{0}, n \in \mathbb{N}^\dagger$.

\begin{proof}
	Clearly, $ \mathbb{S}_1 \subseteq \mathbb{S}_0$ because $ \mathbb{S}_1$ contains additional constraints on $x$. 
	With the condition that $\mathbb{S}_{n} \subseteq \mathbb{S}_{n-1}$, we have 
	\begin{align}
		\begin{aligned}
		\mathbb{S}_{n+1} &=         \{ x | x \in \mathbb{S}_0,  A_n^{cl} x \oplus \mathbb{W} \oplus \gamma_0 \mathbb{P}_\Delta \mathbb{Z}_m \subseteq \mathbb{S}_{n} \}  \\
		&\subseteq \{ x | x \in \mathbb{S}_0,  A_n^{cl} x \oplus \mathbb{W} \oplus \gamma_0 \mathbb{P}_\Delta \mathbb{Z}_m \subseteq \mathbb{S}_{n-1} \}  \\
		&= \mathbb{S}_{n}. 
		\end{aligned} 
	\end{align}
	Thus, property 1 holds.
\end{proof}

Then, $\mathbb{S}_\infty$ can be determined recursively. 
As concluded in \cite{kolmanovsky1998theory}, when $A_n^{cl}$ is Schur, $S_\infty$ can be determined in finite steps as $\mathbb{S}_\infty=\mathbb{S}_{n^*}$. 
For all $n \geq n^*, \mathbb{S}_{n+1}=\mathbb{S}_{n}$, which is employed as the terminal condition of the recursive algorithm.

Notate $\gamma_n=\min_\gamma \{ \gamma |\mathbb{S}_n\subseteq\gamma\mathbb{X}_m \}, n\in\mathbb{N}^\dagger$. 
According to property 1, $\gamma_n \leq \gamma_0$.
If $\gamma_n < \gamma_0$, which means $\mathbb{S}_n \subseteq \gamma_n \mathbb{X}_m$, $\gamma_0$ should be updated by $\gamma_n$ to reduce the size of $\gamma_0 \mathbb{P}_\Delta \mathbb{Z}_m$, and consequently, reduce conservativeness.
Then, the recursive algorithm to obtain $\mathbb{S}_\infty$ for systems subject to both AD and MD is concluded in Algorithm 1.

\begin{algorithm}[H]
	\label{alg:Sinfty}
	\caption{Algorithm to obtain output admissible set $\mathbb{S}_\infty$.}
	\begin{algorithmic}
		\STATE \textbf{Input:} $\mathbb{X}_m, \mathbb{X}_{xu}, \mathbb{W}, \mathbb{P}_\Delta \mathbb{Z}_m , A_n^{cl}$, and  $\gamma_0$
		\STATE \textbf{Output:} $\mathbb{S}_\infty$
		
		\STATE $n=-1$;
		\REPEAT
		\STATE $n = n+1$;
		\IF {$n=0$}
		\STATE $\mathbb{S}_0=\gamma_0\mathbb{X}_m\cap\mathbb{X}_{xu}$;
		\ELSE
		\STATE obtain $\mathbb{S}_n$ according to (\ref{eqn:Srecursive});
		\STATE obtain $\gamma_n=\min_\gamma \{ \gamma |\mathbb{S}_n\subseteq\gamma\mathbb{X}_m \}$;
		\IF {$\gamma_n<\gamma_0$}
		\STATE  $\gamma_0=\gamma_n$;
		\STATE $n=-1$;
		\ENDIF		
		\ENDIF	
		\UNTIL{$\mathbb{S}_n=\mathbb{S}_{n-1}$};
		\STATE{$\mathbb{S}_\infty=\mathbb{S}_{n}$};
		\RETURN $\mathbb{S}_\infty$.
	\end{algorithmic}
\end{algorithm}

\textbf{Remark 3:}
It can be seen that $\mathbb{S}_\infty$ is related to $\gamma_0$. 
With $\mathbb{S}_0=\gamma_0\mathbb{X}_m\cap\mathbb{X}_{xu}$, the corresponding terminal set is notated as $\mathbb{S}_\infty (\gamma _0)$.
Notate $\gamma_\infty = \min_\gamma \{ \gamma | \mathbb{S}_\infty(\gamma_0) \in \gamma_\infty \mathbb{X}_m\} $, then $\mathbb{S}_\infty(\gamma_\infty)=\mathbb{S}_\infty(\gamma _0)$ and $\gamma_\infty \leq \gamma_0$.
In the following parts, $\mathbb{S}_\infty(\gamma_0)$ refers to those sets with $\gamma_\infty=\gamma_0$.

\textbf{Remark 4:} A larger $\gamma_0$ cannot guarantee a $\mathbb{S}_\infty$ with a larger volume because of the expansion of $\gamma_0 \mathbb{P}_\Delta \mathbb{Z}_m$.
Generally,  $\mathbb{S}_\infty(\gamma_\infty^1) \supseteq \mathbb{S}_\infty(\gamma_\infty^1)$ does not hold even when $\gamma_\infty^1 > \gamma_\infty^2$. 

It is found that if $\gamma_0 > \bar{\gamma}$ or $\gamma_0 < \underline{\gamma}$, the corresponding $\mathbb{S}_\infty$ is empty, where
\begin{align} 
	\begin{aligned}
		\bar{\gamma}       &= \min \{ \bar{\gamma}_1,\bar{\gamma}_2 \} ,\\
		\bar{\gamma}_1     &= \min_\gamma \{ \gamma | \mathbb{X}_{xu}\subseteq\gamma\mathbb{X}_{m} \},\\
		\bar{\gamma}_2     &= \max_\gamma \{ \gamma | \mathbb{W} \oplus \gamma \mathbb{P}_\Delta \mathbb{Z}_m \subseteq \mathbb{X}_{xu} \}, \\
		\underline{\gamma} &= \min_\gamma \{ \gamma | \mathbb{W} \oplus \gamma \mathbb{P}_\Delta \mathbb{Z}_m \subseteq \gamma\mathbb{X}_m\} .
	\end{aligned}
\end{align}

Thus,  $\gamma_0$ should be chosen in $\mathbb{R}_{[\underline{\gamma}, \bar{\gamma}]}$. 
It should be pointed out that $\gamma_0 \in \mathbb{R}_{[\underline{\gamma}, \bar{\gamma}]}$ is only a necessary condition that Algorithm 1 returns a non-empty $\mathbb{S}_\infty$.

\subsection{RMPC Controller}
Similar as \cite{chisci2001systems}, the cost function is chosen to be 
\begin{align} \label{eqn:costFunction}
	J(\bm{v}(t), \bm{\lambda}(t))=\sum_{i=0}^{N-1}\|v(k|t)\|_\psi^2
\end{align}
where $\psi$ is positive definite. The quadratic optimization problem $\mathbb{QP}(x(t))$ to be solved online is given as 
\begin{align} 
	\label{eqn:QP}
	\begin{aligned}
		\min_{\bm{v}(t),\bm{\lambda}(t)}   \quad & J(\bm{v}(t), \bm{\lambda}(t)) \\
		s.t.               \quad & \text{(\ref{eqn:QP_constraints}), (\ref{eqn:QP_constraints_equality})}.
	\end{aligned}
\end{align}

The proposed CC-VT RMPC controller is given as
\begin{align}  \label{eqn:RMPCcontrolLaw}
	\pi(x(t))=Kx(t)+v^*(0|t).
\end{align}

With all the above analysis, the proposed CC-VT RMPC method is concluded in Algorithm 2.

\begin{algorithm}[H] 	\caption{Proposed CC-VT RMPC method.}
	\begin{algorithmic}
		\STATE \textbf{Input:} $\mathbb{Z},\mathbb{P}_\Delta,\mathbb{W}, A_n,B_n$ and $K$ which satisfy (A1) to (A4), polytope $\mathbb{Z}_m$, $\mathbb{S}_\infty$ satisfying (A5), and matrix $\psi \succ 0$;

		\STATE \textbf{Online:}
		\STATE Step 1: Obtain current state $x(t)$;
		\STATE Step 2: solve optimization problem (\ref{eqn:QP});
		\STATE Step 3: obtain control input $u(t)$ according to (\ref{eqn:RMPCcontrolLaw});
		\STATE Step 4: control the system (\ref{eqn:actualSystem});
		\STATE Step 5: $t=t+1$ and go to step 1.
	\end{algorithmic}
\end{algorithm}

The robust constraint satisfaction, robust recursive feasibility, and robust stability properties of the proposed method are concluded in Theory 1.

\textbf{Theory 1.}
 For system (\ref{eqn:actualSystem}) with the proposed CC-VT RMPC method, if optimization problem (\ref{eqn:QP}) is feasible at $t$, then
 
 (i)(robust constraint satisfaction) constraint (\ref{eqn:systemConstraints}) is satisfied;
 
 (ii)(robust recursive feasibility) optimization problem (\ref{eqn:QP}) remains feasible at $t+1$; 
 
 (iii)(robust stability) state $x(t)$ converges to a neighborhood of the origin.
 
 \begin{proof}
 	See Appendix A.
 \end{proof}
 
 Suppose $x(\infty) \in \lambda^{\infty} \mathbb{X}_m$, then $w_\mathrm{M}(\infty) \in \lambda^{\infty}  \mathbb{P}_\Delta \mathbb{Z}_m$. 
 According to \cite{chisci2001systems}, $x(t)$ converges to $\sum_{i=0}^{\infty} (A_n^{cl})^i (\mathbb{W} \oplus \lambda^{\infty}  \mathbb{P}_\Delta \mathbb{Z}_m)$, 
 that is, $\sum_{i=0}^{\infty} (A_n^{cl})^i \mathbb{W} \oplus \lambda^{\infty} \sum_{i=0}^{\infty} \mathbb{P}_\Delta \mathbb{Z}_m$,
 which is known as "the minimal disturbance invariant set".
 Here, $\sum_{i=0}^{\infty} (A_n^{cl})^i \mathbb{W}$ and $\sum_{i=0}^{\infty} \mathbb{P}_\Delta \mathbb{Z}_m$ is infinitely determined 
 but can be approximated according to \cite{rakovic2005invariant} and \cite{ong2006minimal}.
 Then, $\lambda^{\infty}$ can be determined by the following optimization problem.
 \begin{align}  
 	\label{eqn:finalStage}
 	\begin{aligned}
 		&\lambda^{\infty} = \min_\lambda   \quad  \lambda \\
 		&s.t.   \quad \sum_{i=0}^{k-1} (A_n^{cl})^i \mathbb{W} \oplus \lambda_{\infty} \sum_{i=0}^{k-1} \mathbb{P}_\Delta \mathbb{Z}_m \subseteq \lambda_{\infty} \mathbb{X}_m.
 	\end{aligned}
 \end{align}
 
  In the optimization problem (\ref{eqn:QP}), $v(k|t)$ gets minimized directly. The following  property shows that although $\lambda(k|t)$ is not included in the cost function, 
  $\lambda(k|t)$ gets minimized when it is necessary.
 
 \textbf{Property 2.} 
  Notate the optimal solution to optimization problem (\ref{eqn:QP}) as 
 \begin{align}
 	{\bm{v}}^*({t}) &=[{v}^{*'}(0|t),{v}^{*'}(1|t),...,{v}^{*'}(N-1|t)]', \\
 	\bm{\lambda}^*(t) &=[\lambda^*(0|t),\lambda^*(1|t),...,\lambda^*(N-1|t)]'. 
 \end{align}
 With $\bar{x}^*(0|t) = x(t)$,  $\bar{x}^*(k|t), k\in \mathbb{N}_{[1,N]}$ are obtained according to (\ref{eqn:dynamicConstraint}).
 Notate $\bm{\bar{x}}^*({t}) = [\bar{x}^{*'}(0|t),\bar{x}^{*'}(1|t),...,\bar{x}^{*'}(N|t)]'$.
 
 Then, the solution with ${\bm{v}(t)} ={\bm{v}}^*({t})$ and 
 						 $\bm{\lambda}(t) =\bm{\lambda}^\dagger(t)=[\lambda^\dagger(0|t),\lambda^\dagger(1|t),...,\lambda^\dagger(N-1|t)]'$
 is feasible to optimization problem (\ref{eqn:QP}) and yields the same cost, where 
 \begin{align}  	\label{eqn:propertyLambda}
 	\begin{aligned}	
 	& {\lambda}^\dagger(k|t) = \min_{\lambda(k|t)} \lambda(k|t) \\
 	&s.t. ~  \begin{bmatrix}\bar{x}^*(k|t)\\K\bar{x}^*(k|t)+v^*(k|t)\end{bmatrix} \oplus \begin{bmatrix}I\\K\end{bmatrix} \overline{\mathbb{E}}(k|t) \subseteq \lambda(k|t)\mathbb{Z}_m.
 	\end{aligned}
 \end{align}
 
\begin{proof}
	With ${\bm{v}}^*({t})$, $\bm{x}^*({t})$, and $\bm{\lambda}^\dagger(t)$, constraints (\ref{eqn:initialConstraints}) and (\ref{eqn:dynamicConstraint}) are satisfied.
	Clearly, $\lambda^\dagger(k|t) \leq \lambda^*(k|t)$.
	Define 	$\overline{\mathbb{E}}^\dagger (0|t)=0,
	 \overline{\mathbb{E}}^\dagger (k+1|t) = A_n^{cl}\overline{\mathbb{E}}^\dagger(k|t) \oplus \mathbb{W} \oplus \lambda^\dagger(k|t) \mathbb{P}_\Delta \mathbb{Z}_m$.
	 
	With $\overline{\mathbb{E}}^\dagger (k|t) \subseteq \overline{\mathbb{E}} (k|t)$, we have
	\begin{align}
		\begin{aligned}
		\overline{\mathbb{E}}^\dagger (k+1|t)  &= A_n^{cl}\overline{\mathbb{E}}^\dagger(k|t) \oplus \mathbb{W} \oplus \lambda^\dagger(k|t) \mathbb{P}_\Delta \mathbb{Z}_m \\
		                                     &\subseteq 
		A_n^{cl}\overline{\mathbb{E}}(k|t) \oplus \mathbb{W} \oplus \lambda(k|t) \mathbb{P}_\Delta \mathbb{Z}_m \\
		& =\overline{\mathbb{E}} (k+1|t).
		\end{aligned}
	\end{align}
	
	Then, with $\overline{\mathbb{E}}^\dagger (0|t) = \overline{\mathbb{E}} (0|t)$, it can be proved recursively that $\overline{\mathbb{E}}^\dagger (k|t) \subseteq \overline{\mathbb{E}} (k|t), k \in \mathbb{N}_{[0,N]}$. Then, constraints (\ref{eqn:feasibleConstraint}) and (\ref{eqn:terminalConstraint}) imply that
	\begin{align}
		\begin{bmatrix}\bar{x}^*(k|t)\\K\bar{x}^*(k|t)+v^*(k|t)\end{bmatrix} &\in \mathbb{Z}\ominus\begin{bmatrix}I\\K\end{bmatrix}\overline{\mathbb{E}}^\dagger(k|t), \\
		\bar{x}^*(N|t) &\in \mathbb{S}_\infty \ominus \overline{\mathbb{E}}^\dagger(N|t).
	\end{align}
	
	Further, according to (\ref{eqn:propertyLambda}), 
	\begin{align}
		\begin{aligned}
	&\begin{bmatrix}\bar{x}^*(k|t)\\K\bar{x}^*(k|t)+v^*(k|t)\end{bmatrix} \oplus \begin{bmatrix}I\\K\end{bmatrix}\overline{\mathbb{E}}^\dagger (k|t)   \\
	&\subseteq \begin{bmatrix}\bar{x}^*(k|t)\\K\bar{x}^*(k|t)+v^*(k|t)\end{bmatrix} \oplus \begin{bmatrix}I\\K\end{bmatrix}\overline{\mathbb{E}} (k|t)  \\
	&\subseteq \lambda^\dagger(k|t)\mathbb{Z}_m.
		\end{aligned}
	\end{align}
	
	This leads to
	\begin{align}
	\begin{bmatrix}\bar{x}^*(k|t)\\K\bar{x}^*(k|t)+v^*(k|t)\end{bmatrix} \in \lambda^\dagger(k|t)\mathbb{Z}_m \ominus\begin{bmatrix}I\\K\end{bmatrix}\overline{\mathbb{E}}^\dagger(k|t). 
	\end{align}
		
	Thus, constraints in optimization problem (\ref{eqn:QP}) are satisfied with ${\bm{v}}^*({t})$, $\bm{x}^*({t})$, and $\bm{\lambda}^\dagger(t)$.
	Meanwhile, $J(\bm{v}^*(t), \bm{\lambda}^\dagger(t)) = J(\bm{v}^*(t), \bm{\lambda}^*(t))$.
	Thus, property 2 holds.	
\end{proof} 

The optimization problems (\ref{eqn:QP}) is computationally tractable and there are no need to reformulate (\ref{eqn:QP}) except updating the initial constraint (\ref{eqn:initialConstraints}) with the current state $x(t)$. 
With the following Lemma, it is easy to obtain the linear-constraint expressions of  (\ref{eqn:addFeasibleConstraint}), (\ref{eqn:feasibleConstraint}), (\ref{eqn:terminalConstraint}) for implementation. 

\textbf{Lemma 1.}
(\cite{blanchini2008set}, Chapter 3 )
For convex set $\mathbb{X}_1{:}\{x\in\mathbb{R}^n|H_{\mathbb{X}_1} x\leq h_{\mathbb{X}_1}\},H_{\mathbb{X}_1} \in\mathbb{R}^{N_{\mathbb{X}_1}^{\mathrm{lcon}}\times n},h_{\mathbb{X}_1} \in\mathbb{R}^{N_{\mathbb{X}_1}^{\mathrm{lcon}}}$ and convex set $\mathbb{X}_2$,

(i) $\mathbb{X}_1\ominus\mathbb{X}_2 = \begin{Bmatrix}x|H_{\mathbb{X}_1} x\leq h_{\mathbb{X}_1}-\Delta_{\mathbb{X}_2}^{\mathbb{X}_1}\end{Bmatrix}$, 
where $\Delta_{\mathbb{X}_2}^{\mathbb{X}_1}\in\mathbb{R}^{N_{\mathbb{X}_{1}}^{\mathrm{lcon}}}$ and the $i$th element of  $\Delta_{\mathbb{X}_2}^{\mathbb{X}_1}$ is determined by
\begin{align}
	\begin{aligned}
	\begin{bmatrix}\Delta_{\mathbb{X}_2}^{\mathbb{X}_1}\end{bmatrix}_i = &\max_x ~  [H_{\mathbb{X}_1}]_i x \\
	s.t. ~ &x\in\mathbb{X}_2.
	\end{aligned}
\end{align}

(ii) $\forall\lambda_1\geq0,\lambda_2\geq0$
\begin{align}
	\lambda_1\mathbb{X}_1\ominus\lambda_2\mathbb{X}_2=\{x|C_1x\leq\lambda_1b_1-\lambda_2 \Delta_{\mathbb{X}_2}^{\mathbb{X}_1}\}.
\end{align}

$N_{\mathbb{X}_1}$ represents the number of linear constraints of $\mathbb{X}_1$. In the following parts, the linear-constraint expression and the number of linear constraints of a polytope are notated in a similar way.

With (\ref{eqn:E_0}), (\ref{eqn:E_E_W_WP}), and (\ref{eqn:WP_kt}), it is obtained that 
\begin{align}  \label{eqn:ebar}
	\begin{aligned}
		  \overline{\mathbb{E}}(k|t) &= \sum_{i=0}^{k-1}(A_n^{cl})^i\mathbb{W} \\
		& \oplus\sum_{i=0}^{k-1}\lambda(k-1-i|t)(A_n^{cl})^i \mathbb{P}_\Delta \mathbb{Z}_m, 
		k\in\mathbb{Z}_{[1,N]}.
	\end{aligned}
\end{align}

Notate $\mathbb{Z}_m=\{z|H_{\mathbb{Z}_m}z\leq h_{\mathbb{Z}_m}\}$,
$H_{\mathbb{Z}_m}\in\mathbb{R}^{N_{\mathbb{Z}_m}^{\mathrm{lcon}}\times(n+m)}$,
$h_{\mathbb{Z}_m}\in\mathbb{R}^{N_{\mathbb{Z}_m}^{\mathrm{lcon}}}$.
According to Lemma 1, the linear-constraint expression of  $\lambda(k|t)\mathbb{Z}_m\ominus\begin{bmatrix}I\\K\end{bmatrix}\overline{\mathbb{E}}(k|t)$ is given as
\begin{align} 
	\begin{aligned}
		&\lambda(k|t)\mathbb{Z}_m\ominus\begin{bmatrix}I\\K\end{bmatrix}\overline{\mathbb{E}}(k|t) \\
		&=\left\{z|H_{\mathbb{Z}_m}z + \sum_{i=0}^{k-1}\lambda(k-1-i|t)\Delta_{\begin{bmatrix}I\\K\end{bmatrix} (A_n^{cl})^i \mathbb{P}_\Delta \mathbb{Z}_m} ^ {\mathbb{Z}_m}  \right.\\
		& ~~~~~~~~~~~~~~~~ \left. -\lambda(k|t) h_{\mathbb{Z}_m}  \leq - \Delta_{\begin{bmatrix}I\\K\end{bmatrix}\sum_{i=0}^{k-1}(A_n^{cl})^i\mathbb{W}}^{\mathbb{Z}_m} \right\} .
	\end{aligned}
\end{align}

The linear-constraint expression of 
(\ref{eqn:feasibleConstraint}) and (\ref{eqn:terminalConstraint}) can be obtained in the same way. 
All those elements in the linear-constraint expressions can be obtained offline and no additional online computations are needed.

\textbf{Remark 5:}
The quadratic optimization problem to be solve online is expressed in the form of (\ref{eqn:QP}) for the ease of illustration. 
For implementation, equality (\ref{eqn:QP_constraints_equality}) should be integrated  with constraints (\ref{eqn:QP_constraints}). 
According to the analysis above, equality constraints (\ref{eqn:E_0}), (\ref{eqn:E_E_W_WP}), and (\ref{eqn:WP_kt}) are integrated with constraints (\ref{eqn:QP_constraints}).
(\ref{eqn:uBar}) is integrated with constraints (\ref{eqn:QP_constraints}) by substituting (\ref{eqn:uBar}) into constraints (\ref{eqn:dynamicConstraint}), (\ref{eqn:addFeasibleConstraint}) and (\ref{eqn:feasibleConstraint}).
Alternatively, equality constraints (\ref{eqn:initialConstraints}) and (\ref{eqn:dynamicConstraint}) can be used to get the expression of $\bar{x}(k|t)$ and then be integrated with (\ref{eqn:addFeasibleConstraint}), (\ref{eqn:feasibleConstraint}), and (\ref{eqn:terminalConstraint}).
This reduces the number of decision variables and shorten the time required to solve the corresponding quadratic optimization problem.

\section{Optimization of Container} \label{sec:optimizationContainerSet}
In this section, container $\mathbb{Z}_m$ is optimized. 
Although $\mathbb{Z}_m$ can be chosen to be any polytope containing the origin in its interior, 
method to find a "better" $\mathbb{Z}_m$ is proposed in this section to make the CC-VT RMPC method less conservative. 

Compared to a container $\mathbb{Z}_m^1$, $\mathbb{Z}_m^2$ is better when the region of attraction of the optimization problem (\ref{eqn:QP}) with $\mathbb{Z}_m = \mathbb{Z}_m^2$ is larger than that of (\ref{eqn:QP}) with  $\mathbb{Z}_m = \mathbb{Z}_m^1$, 
where the attraction region is defined as the set af $x(t)$ which makes optimization problem (\ref{eqn:QP}) feasible.

It should be noted that $\mathbb{Z}_m^2 = \alpha \mathbb{Z}_m^1,\alpha>0$ is not better than  $\mathbb{Z}_m^1$. 
If the solution with ${\bm{v}(t)}$ and $\bm{\lambda}(t)$ is feasible to optimization problem (\ref{eqn:QP}) with $\mathbb{Z}_m = \mathbb{Z}_m^2$, 
then, the solution with ${\bm{v}(t)}$ and $\alpha\bm{\lambda}(t)$ is feasible to (\ref{eqn:QP}) with $\mathbb{Z}_m = \mathbb{Z}_m^1$.
This means, the size of $\mathbb{Z}_m$ make no sense and it is the shape of it that plays the significant role. 

\subsection{Optimization of $\mathbb{Z}_m^0$} 

Given a container $\mathbb{Z}_m^0$, the vertex expression of $\mathbb{P}_\Delta \mathbb{Z}_m^0$ can be determined according to the following lemma. 

\textbf{Lemma 2:} Given polytopes 
\begin{align*}
	\mathbb{P}_\Delta &= \mathbb{CH}\{ \Delta P_i ^{\mathrm{vert}}\} \subset \mathbb{R}^{n \times (n+m)},  i\in\mathbb{N}_{\left[1,N_{\mathbb{P}_\Delta}^{\mathrm{vert}} \right]} \\
	\mathbb{Z}_m^0    &= \mathbb{CH}\{z_j^{\mathrm{vert}}\} \subset \mathbb{R}^{n+m},j \in \mathbb{N}_{[1,N_{\mathbb{Z}_m^0}^{\mathrm{vert}}]}
\end{align*}
$\mathbb{P}_\Delta \mathbb{Z}_m^0$ is a polytope which can be expressed by
\begin{align}
	\mathbb{P}_\Delta \mathbb{Z}_m^0 = \mathbb{CH} \{ \Delta P_i^{\mathrm{vert}} z_j^{\mathrm{vert}} \}.
\end{align}

\begin{proof}
	By definition, $\mathbb{P}_\Delta \mathbb{Z}_m^0 = \{ P_\Delta z | P_\Delta \in \mathbb{P}_\Delta, z \in \mathbb{Z}_m^0\}$.
	Then, for all $w_\mathrm{M} \in \mathbb{P}_\Delta \mathbb{Z}_m^0$, there exist $P_\Delta^0 \in \mathbb{P}_\Delta, z^0 \in \mathbb{Z}_m^0$ 
	such that $w_\mathrm{M}=P_\Delta^0 z^0$.
	
	Meanwhile, $P_\Delta^0$ and $z^0$ can be expressed as 
	\begin{align}
		P_\Delta^0 = \sum_{i=1}^{ N_{\mathbb{P}_\Delta}^{\mathrm{vert}}}  \alpha_i \Delta P_i^{\mathrm{vert}}, 
		&\sum_{i=1}^{ N_{\mathbb{P}_\Delta}^{\mathrm{vert}}}  \alpha_i = 1, \alpha_i \geq 0, \\
		z^0 = \sum_{j=1}^{ N_{\mathbb{Z}_m^0}^{\mathrm{vert}} } \beta_j z_j ^{\mathrm{vert}}, 
		&\sum_{j=1}^{ N_{\mathbb{Z}_m^0}^{\mathrm{vert}} } \beta_j = 1, \beta_j \geq 0.
	\end{align}
	
	Then, 
	\begin{align}
		w_\mathrm{M} = P_\Delta^0 z^0 &= \sum_{i=1}^{ N_{\mathbb{P}_\Delta}^{\mathrm{vert}}} \sum_{j=1}^{ N_{\mathbb{Z}_m^0}^{\mathrm{vert}}} 
		\alpha_i \beta_j \Delta P_i^{\mathrm{vert}} z_j ^{\mathrm{vert}},  \\
		\sum_{i=1}^{ N_{\mathbb{P}_\Delta}^{\mathrm{vert}}} \sum_{j=1}^{ N_{\mathbb{Z}_m^0}^{\mathrm{vert}}} \alpha_i \beta_j 
		&= \sum_{i=1}^{ N_{\mathbb{P}_\Delta}^{\mathrm{vert}}} (\alpha_i \sum_{j=1}^{ N_{\mathbb{Z}_m^0}^{\mathrm{vert}}} \beta_j)
		= \sum_{i=1}^{ N_{\mathbb{P}_\Delta}^{\mathrm{vert}}} \alpha_i 
		= 1.
	\end{align}					 	  
	Thus, $z^0$ lies in the convex hull of $\Delta P_i^{\mathrm{vert}} z_j^{\mathrm{vert}}, 
	i\in\mathbb{N}_{\left[1,N_{\mathbb{P}_\Delta}^{\mathrm{vert}} \right]}, 
	j \in \mathbb{N}_{[1,N_{\mathbb{Z}_m^0}^{\mathrm{vert}}]}$ and Lemma 2 holds.
\end{proof}

Then, with the vertex-expression of $\mathbb{P}_\Delta \mathbb{Z}_m^0$, the equivalent linear-constraint expression of $\mathbb{P}_\Delta \mathbb{Z}_m^0$ can be obtained, which is notated as
\begin{align} 
	\begin{aligned}
		\mathbb{P}_\Delta \mathbb{Z}_m^0 = \{ w|H_{\mathbb{P}_\Delta \mathbb{Z}_m^0} w \leq h_{\mathbb{P}_\Delta \mathbb{Z}_m^0} \},
	\end{aligned}
\end{align}
with $H_{\mathbb{P}_\Delta \mathbb{Z}_m^0} \in \mathbb{R}^{N_{\mathbb{P}_\Delta \mathbb{Z}_m^0}^{\mathrm{lcon}} \times n}, 
h_{\mathbb{P}_\Delta \mathbb{Z}_m^0} \in \mathbb{R}^{N_{\mathbb{P}_\Delta \mathbb{Z}_m^0}^{\mathrm{lcon}} }$.

$\mathbb{Z}_m$ is then determined as the set of admissible $z$ which satisfies $\mathbb{P}_\Delta z \subseteq  \mathbb{P}_\Delta \mathbb{Z}_m^0$, 
expressed as 
\begin{align} \label{eqn:expressionZm}
	\mathbb{Z}_m = \{ z | \mathbb{P}_\Delta z \subseteq  \mathbb{P}_\Delta \mathbb{Z}_m^0 \}.
\end{align}
Since $\mathbb{P}_\Delta = \mathbb{CH}\{ \Delta P_i ^{\mathrm{vert}}\}$, then $\mathbb{P}_\Delta z \subseteq  \mathbb{P}_\Delta \mathbb{Z}_m^0$ is equivalent to 
$\Delta P_i ^{\mathrm{vert}} z \in  \mathbb{P}_\Delta \mathbb{Z}_m^0, \forall i\in\mathbb{N}_{\left[1,N_{\mathbb{P}_\Delta}^{\mathrm{vert}} \right]}$.
Thus, the linear-constraint expression of $\mathbb{Z}_m$ is expressed as
\begin{align} \label{eqn:Zm2}
	\begin{aligned}
		\mathbb{Z}_m=\left\{ z |
		H_{\mathbb{P}_\Delta \mathbb{Z}_m^0} \Delta P_i ^{\mathrm{vert}} z \leq h_{\mathbb{P}_\Delta \mathbb{Z}_m^0} ,
		i\in\mathbb{N}_{[1,N_{\mathbb{P}_\Delta}^{\mathrm{vert}}]} \right\}.
	\end{aligned}
\end{align}

\textbf{Theory 2:}
For a given polytope $\mathbb{Z}_m^1$ containing the origin, $\mathbb{Z}_m^2$ is defined by 
$\mathbb{Z}_m^2 = \{ z | \mathbb{P}_\Delta z \subseteq  \mathbb{P}_\Delta \mathbb{Z}_m^1 \}$. 
Then, $\mathbb{Z}_m^1 \subseteq \mathbb{Z}_m^2$.
Further, if $\mathbb{Z}_m^1 \subset \mathbb{Z}_m^2$, then $\mathbb{Z}_m^2$ is better than $\mathbb{Z}_m^1$.

\begin{proof}
 See Appendix B.
\end{proof}

\textbf{Remark 6:}
By default, $\mathbb{Z}_m^0 \subset \mathbb{R}^{n+m}$  can be chosen as an inner polytope of $\mathbb{B}^{n+m}$ to make it representative.
Alternatively, $\mathbb{X}_m^0 \subseteq \mathbb{R}^{n}$ can be chosen to be an inner polytope of  $\mathbb{B}^{n}$, 
and then  $\mathbb{Z}_m^0$ is determined by $\mathbb{Z}_m^0 = \Pi \mathbb{X}_m^0$ where
\begin{align}
	\Pi = \begin{bmatrix}I\\K\end{bmatrix}.
\end{align}

\subsection{Optimization of $\mathbb{PZ}_m^0$} \label{sec:optPZ0}
Alternatively, $\mathbb{P}_\Delta \mathbb{Z}_m^0$, notated as $\overline{\mathbb{W}}_\mathrm{M}^0$, can be specified directly rather than being determined by the product of $\mathbb{P}_\Delta$ and a specified $\mathbb{Z}_m^0$. 
Given $\overline{\mathbb{W}}_\mathrm{M}^0$, $\mathbb{Z}_m^0$ can be determined by 
	$\mathbb{Z}_m^0 = \{ z | \mathbb{P}_\Delta z \subseteq \overline{\mathbb{W}}_\mathrm{M}^0\}$.
$\mathbb{Z}_m^0$ is the maximal set ensuring $\mathbb{P}_\Delta \mathbb{Z}_m^0 \subseteq \overline{\mathbb{W}}_\mathrm{M}^0$.
Then, to enlarge the attraction region, we try to find a "better" $\overline{\mathbb{W}}_\mathrm{M}^0$, notated as $\overline{\mathbb{W}}_\mathrm{M}^\dagger$.
Similarly, given $\overline{\mathbb{W}}_\mathrm{M}^\dagger$, $\mathbb{Z}_m^\dagger$ can be determined by 
	$\mathbb{Z}_m^\dagger = \{ z | \mathbb{P}_\Delta z \subseteq \overline{\mathbb{W}}_\mathrm{M}^\dagger\}$. 

Compare to $\overline{\mathbb{W}}_\mathrm{M}^0$, $\overline{\mathbb{W}}_\mathrm{M}^\dagger$ is better when
 $\overline{\mathbb{W}}_\mathrm{M}^0 \subset \overline{\mathbb{W}}_\mathrm{M}^\dagger$
 and the following two conditions hold.

(i) $\mathbb{Z} \ominus \Pi \overline{\mathbb{E}}_1(k|t) = \mathbb{Z} \ominus \Pi \overline{\mathbb{E}}_2(k|t)$ and 
   $\mathbb{Z}_m^0 \ominus \Pi \overline{\mathbb{E}}_1(k|t) = \mathbb{Z}_m^0 \ominus \overline{\mathbb{E}}_2(k|t)$ 
	where $\overline{\mathbb{E}}_1(0|t)=\overline{\mathbb{E}}_2(0|t)=\{ 0\}$ and 
	\begin{align}  \label{eqn:fbar}
		\begin{aligned}
			\overline{\mathbb{E}}_1(k|t) &= \sum_{i=0}^{k-1}(A_n^{cl})^i\mathbb{W} \\
			& \oplus\sum_{i=0}^{k-1}\lambda(k-1-i|t)(A_n^{cl})^i \overline{\mathbb{W}}_\mathrm{M}^0, \\
			\overline{\mathbb{E}}_2(k|t) &= \sum_{i=0}^{k-1}(A_n^{cl})^i\mathbb{W} \\
			& \oplus\sum_{i=0}^{k-1}\lambda(k-1-i|t)(A_n^{cl})^i \overline{\mathbb{W}}_\mathrm{M}^\dagger, 
			k\in\mathbb{Z}_{[1,N]},
		\end{aligned}
	\end{align}
	
(ii) $\mathbb{S}_\infty^0 \subseteq \mathbb{S}_\infty^\dagger, 
	 \mathbb{S}_\infty^0 \ominus \overline{\mathbb{E}}_1(N|t) = \mathbb{S}_\infty^0 \ominus \overline{\mathbb{E}}_2(N|t)$, 
	 where $\mathbb{S}_\infty^0$ and $\mathbb{S}_\infty^\dagger$ are determined recursively by
\begin{align} \label{eqn:Sdagger}
	\begin{aligned}
	\mathbb{S}_0^0          &= \gamma_\infty \mathbb{X}_m^0 \cap \mathbb{X}_{xu},  \\
	\mathbb{S}_0^\dagger    &= \gamma_\infty \mathbb{X}_m^\dagger\cap\mathbb{X}_{xu}, \\
	\mathbb{S}_n^0    &= \{ x | x \in \mathbb{S}_0^0,  
	A_n^{cl} x \oplus \mathbb{W} \oplus \gamma_\infty \overline{\mathbb{W}}_\mathrm{M}^0 \subseteq \mathbb{S}_{n-1}^0\}, \\
	\mathbb{S}_n^\dagger    &= \{ x | x \in \mathbb{S}_0^\dagger,  
	A_n^{cl} x \oplus \mathbb{W} \oplus \gamma_\infty \overline{\mathbb{W}}_\mathrm{M}^\dagger \subseteq \mathbb{S}_{n-1}^\dagger\}.
	\end{aligned}
\end{align}

\textbf{Theory 3:}
If $\overline{\mathbb{W}}_\mathrm{M}^\dagger$ is better than $\overline{\mathbb{W}}_\mathrm{M}^0$, $\mathbb{Z}_m^\dagger $ is better than $\mathbb{Z}_m$.
	
\begin{proof}
	Since $\overline{\mathbb{W}}_\mathrm{M}^0 \subset \overline{\mathbb{W}}_\mathrm{M}^\dagger$, $\mathbb{Z}_m^0 \subset \mathbb{Z}_m^\dagger$.
	
	According to Lemma 1, 
	it is easy to concluded that for convex sets $\mathbb{X}, \mathbb{X}_1$, and $\mathbb{X}_2$,
	if $\mathbb{X} \ominus \mathbb{X}_1 = \mathbb{X} \ominus \mathbb{X}_2$,
	then $\alpha \mathbb{X} \ominus \beta \mathbb{X}_1 = \alpha \mathbb{X} \ominus \beta \mathbb{X}_2, \alpha>0, \beta >0$.
	
	With $\mathbb{Z}_m^0 \subset \mathbb{Z}_m^\dagger$,
	$\mathbb{Z}_m^0 \ominus \Pi \overline{\mathbb{E}}_1(k|t) = \mathbb{Z}_m^0 \ominus \overline{\mathbb{E}}_2(k|t)$  implies
	$\lambda \mathbb{Z}_m^1 \ominus \Pi \overline{\mathbb{E}}_1(k|t) 
	= \lambda \mathbb{Z}_m^1 \ominus \Pi \overline{\mathbb{E}}_2(k|t) 
	\subset \lambda \mathbb{Z}_m^2 \ominus \Pi \overline{\mathbb{E}}_2(k|t)$.
	
	With $\mathbb{S}_\infty^0 \subseteq \mathbb{S}_\infty^\dagger, 
	\mathbb{S}_\infty^0 \ominus \overline{\mathbb{E}}_1(N|t) = \mathbb{S}_\infty^0 \ominus \overline{\mathbb{E}}_2(N|t)$ implies 
	$\mathbb{S}_\infty^0 \ominus \overline{\mathbb{E}}_1(N|t) \subseteq \mathbb{S}_\infty^\dagger \ominus \overline{\mathbb{E}}_2(N|t)$.
	
	Meanwhile, $\mathbb{Z} \ominus \Pi \overline{\mathbb{E}}_1(k|t) = \mathbb{Z} \ominus \Pi \overline{\mathbb{E}}_2(k|t)$, thus, all feasible solutions to optimization problem (\ref{eqn:QP}) with $\mathbb{Z}_m = \mathbb{Z}_m^0$ are feasible to (\ref{eqn:QP}) with $\mathbb{Z}_m = \mathbb{Z}_m^\dagger$.
	
	On the contrary, since $\mathbb{Z}_m^0 \subset \mathbb{Z}_m^\dagger$, 
	there exists solutions such that constraint (\ref{eqn:addFeasibleConstraint}) is satisfied with $\mathbb{Z}_m = \mathbb{Z}_m^\dagger$ but is not satisfied with $\mathbb{Z}_m = \mathbb{Z}_m^0$.
	
	Thus, if $\overline{\mathbb{W}}_\mathrm{M}^\dagger$ is better than $\overline{\mathbb{W}}_\mathrm{M}^0$, $\mathbb{Z}_m^\dagger $ is better than $\mathbb{Z}_m$.
\end{proof}

\textbf{Lemma 3:}
Suppose $\overline{\mathbb{W}}_\mathrm{M}^0 \subseteq \overline{\mathbb{W}}_\mathrm{M}^\dagger$, 
if $\mathbb{S}_n^0 \ominus \overline{\mathbb{W}}_\mathrm{M}^0 
  = \mathbb{S}_n^0 \ominus \overline{\mathbb{W}}_\mathrm{M}^\dagger,
n \in \mathbb{R}_{[0,n^\dagger]}$ holds, then, 
(i)  $\mathbb{S}_n^0 \subseteq \mathbb{S}_n^\dagger$, 
(ii) $\mathbb{S}_\infty^0 \subseteq \mathbb{S}_\infty^\dagger$ if $\mathbb{S}^0_{n^\dagger} = \mathbb{S}^0_{n^\dagger -1}$ and $\mathbb{S}_{n^\dagger}^\dagger = \mathbb{S}_{n^\dagger -1}^\dagger$.

\begin{proof}
	Since $\overline{\mathbb{W}}_\mathrm{M}^0 \subseteq \overline{\mathbb{W}}_\mathrm{M}^\dagger$, $\mathbb{Z}_m^0 \subseteq \mathbb{Z}_m^\dagger$.
	Then, $\mathbb{S}_n^0 \subseteq \mathbb{S}_n^\dagger$.
	
	With  $\mathbb{S}_{n-1}^0 \subseteq {\mathbb{S}}_{n-1}^\dagger$, 
	$\forall x \in \mathbb{S}_n^0$, we have $x \in \mathbb{S}_0^0 \subseteq {\mathbb{S}}_{0}^\dagger $ and 
	\begin{align}
		\begin{aligned}
			A_n^{cl} x 
			&\in \mathbb{S}_{n-1}^0 \ominus \mathbb{W} \ominus \gamma_\infty \overline{\mathbb{W}}_\mathrm{M}^0 \\
			&= \mathbb{S}_{n-1}^0 \ominus \mathbb{W} \ominus \gamma_\infty \overline{\mathbb{W}}_\mathrm{M}^\dagger \\
			&\subseteq {\mathbb{S}}^\dagger_{n-1} \ominus \mathbb{W} \ominus \gamma_\infty \overline{\mathbb{W}}_\mathrm{M}^\dagger.
		\end{aligned}
	\end{align}
	
	Thus, $\mathbb{S}_{n}^0 \subseteq {\mathbb{S}}^\dagger_{n}$.
	Meanwhile, with $\overline{\mathbb{S}}_{0} \subseteq \mathbb{S}^\dagger_{0}$, we have
	$\overline{\mathbb{S}}_{n} \subseteq \mathbb{S}^\dagger_{n}, i \in \mathbb{R}_{[0,n^\dagger]}$ and (i) holds.
	
	$\mathbb{S}_{n^\dagger} = \mathbb{S}_{n^\dagger -1}$ and $\mathbb{S}_{n^\dagger}^\dagger = \mathbb{S}_{n^\dagger -1}^\dagger$ imply that $\mathbb{S}_{n^\dagger} = \mathbb{S}_\infty$ and $\mathbb{S}_{n^\dagger}^\dagger = \mathbb{S}_\infty^\dagger$.
	According to (i), $\mathbb{S}_\infty \subseteq \mathbb{S}_\infty^\dagger$ and (ii) holds.
\end{proof}

As shown in (\ref{eqn:ebar}), $\overline{\mathbb{E}}(k|t)$ consists of two parts, one related to $\mathbb{W}$ and the other related to $\overline{\mathbb{W}}_\mathrm{M}^0$ or $\overline{\mathbb{W}}_\mathrm{M}^\dagger$. 
Then, according to Lemma 1, except $\overline{\mathbb{W}}_\mathrm{M}^0 \subset \overline{\mathbb{W}}_\mathrm{M}^\dagger$, 
the other conditions for a better $\overline{\mathbb{W}}_\mathrm{M}^\dagger$ are equivalent to
\begin{align}
	   \Delta ^{ \mathbb{Z} } _{\Pi (A_n^{cl})^i \overline{\mathbb{W}}_\mathrm{M}^0         } 
	&= \Delta ^{ \mathbb{Z} } _{\Pi (A_n^{cl})^i \overline{\mathbb{W}}_\mathrm{M}^\dagger   }, 	      \label{eqn:WMDagger1}\\ 
	   \Delta ^{ \mathbb{Z}^0_m } _{\Pi (A_n^{cl})^i \overline{\mathbb{W}}_\mathrm{M}^0 } 
	&= \Delta ^{ \mathbb{Z}^0_m } _{\Pi (A_n^{cl})^i \overline{\mathbb{W}}_\mathrm{M}^\dagger },      \label{eqn:WMDagger2} \\    
	   \Delta ^{ \mathbb{S}_\infty^0 } _{ (A_n^{cl})^i \overline{\mathbb{W}}_\mathrm{M}^0 } 
	&= \Delta ^{ \mathbb{S}_\infty^0 } _{ (A_n^{cl})^i \overline{\mathbb{W}}_\mathrm{M}^\dagger }, 	  \label{eqn:WMDagger3} \\ 
	   \Delta ^{ \mathbb{S}_j^0 } _{\overline{\mathbb{W}}_\mathrm{M}^0 } 
	&= \Delta ^{ \mathbb{S}_j^0 } _{\overline{\mathbb{W}}_\mathrm{M}^\dagger }, 			          \label{eqn:WMDagger4} 
\end{align}
Where $ i \in \mathbb{Z}_{[1,N]},  j \in \mathbb{Z}_{[1,n^\dagger]}$. 

If the vertex expression of $\overline{\mathbb{W}}_\mathrm{M}^0$ is specified as $\overline{\mathbb{W}}_\mathrm{M}^0 = \mathbb{CH} \{ w^{\mathrm{vert}}_l\}, 
	l \in \mathbb{N}_{[1,N_{\overline{\mathbb{W}}_\mathrm{M}^0}^{\mathrm{vert}}]} $, 
the vertex expression of $\overline{\mathbb{W}}_\mathrm{M}^\dagger$ can be determined as
$\overline{\mathbb{W}}_\mathrm{M}^0 = \mathbb{CH} \{ \beta_l^* w^{\mathrm{vert}}_l\}$,
where $ \beta_l^*$ are obtained from the following optimization problem.
\begin{align} \label{eqn:QP_WMdagger_1}
	\begin{aligned}
		\max_{\beta_1, \beta_2, ..., \beta_{{N}^{\mathrm{vert}}_{\overline{\mathbb{W}}_\mathrm{M}^0} } } ~ &\sum_{l=1}^{{N}^{\mathrm{vert}}_{\overline{\mathbb{W}}_\mathrm{M}^0}} \beta_l \\
		s.t. ~~ & \text{(\ref{eqn:WMDagger1}), (\ref{eqn:WMDagger2}), (\ref{eqn:WMDagger3}), (\ref{eqn:WMDagger4}), }  \\
		 & \beta_l \geq 1, l \in \mathbb{N}_{[1,N_{\overline{\mathbb{W}}_\mathrm{M}^0}^{\mathrm{vert}}]}.
	\end{aligned}
\end{align}

Similarly, if the linear-constraint expression of $\overline{\mathbb{W}}_\mathrm{M}^0$ is specified as 
$\overline{\mathbb{W}}_\mathrm{M}^0 = \{w | H_{\overline{\mathbb{W}}_\mathrm{M}^0} w \leq h_{\overline{\mathbb{W}}_\mathrm{M}^0}\}, 
l \in \mathbb{N}_{[1,N_{\overline{\mathbb{W}}_\mathrm{M}^0}^{\mathrm{lcon}}]}$, 
the linear-constraint expression of $\overline{\mathbb{W}}_\mathrm{M}^\dagger$ can be determined by
$\overline{\mathbb{W}}_\mathrm{M}^0 = \{w | H_{\overline{\mathbb{W}}_\mathrm{M}^0} w \leq h^*_{\overline{\mathbb{W}}_\mathrm{M}^0}\}$,
where $[h^*_{\overline{\mathbb{W}}_\mathrm{M}^0}]_l = \beta_l^* [h_{\overline{\mathbb{W}}_\mathrm{M}^0}]_l $ and $\beta_l^*$ is determined by the optimization problem (\ref{eqn:QP_WMdagger_1}) with ${N}^{\mathrm{vert}}_{\overline{\mathbb{W}}_\mathrm{M}^0}$ replaced by ${N}^{\mathrm{lcon}}_{\overline{\mathbb{W}}_\mathrm{M}^0}$.

\textbf{Remark 7:}
Optimization problem (\ref{eqn:QP_WMdagger_1}) is always feasible with 
$\beta_l = 1, \forall l \in \mathbb{N}_{[1,N_{\overline{\mathbb{W}}_\mathrm{M}^0}^{\mathrm{vert}}]}$ as a feasible solution. In this case, $\overline{\mathbb{W}}_\mathrm{M}^0 = \overline{\mathbb{W}}_\mathrm{M}^\dagger$.
Or else, $\overline{\mathbb{W}}_\mathrm{M}^0 \subset \overline{\mathbb{W}}_\mathrm{M}^\dagger$ and $\overline{\mathbb{W}}_\mathrm{M}^\dagger$ is better than $\overline{\mathbb{W}}_\mathrm{M}^0$.
The choice of cost function of (\ref{eqn:QP_WMdagger_1}) is non-unique, 
and it is possible to allocate different weights to $\beta_l$. 

\textbf{Remark 8:}
Constraints (\ref{eqn:WMDagger1}), (\ref{eqn:WMDagger2}), (\ref{eqn:WMDagger3}), and (\ref{eqn:WMDagger4}) are strict and may easily yield  $\beta_l^* = 1, \forall l \in \mathbb{N}_{[1,N_{\overline{\mathbb{W}}_\mathrm{M}^0}^{\mathrm{vert}}]}$,
and thus, should be relaxed.
Because $A_n^{cl}$ is stable, $(A_n^{cl})^i \overline{\mathbb{W}}_\mathrm{M}^\dagger$ diminishes with the increase of $i$. 
Then, $i \in \mathbb{N}_{[1,N_i]}, j \in \mathbb{N}_{[1,N_j]}$ may be considered instead,
where $N_i < N, N_j < n^\dagger$ are smaller positive integers.
Meanwhile, it is reasonable to drop constraints (\ref{eqn:WMDagger2}), (\ref{eqn:WMDagger3}), and (\ref{eqn:WMDagger4}) 
because $\mathbb{Z}_m^0$ and $\mathbb{S}_\infty^0$ will be updated by $\mathbb{Z}_m^\dagger$ and $\mathbb{S}_\infty^\dagger$ and the constraints (\ref{eqn:addFeasibleConstraint}) and (\ref{eqn:terminalConstraint}) change distinctly. 
With the relaxed constraints, it is not guaranteed that $\overline{\mathbb{W}}_\mathrm{M}^\dagger$ is better than $\overline{\mathbb{W}}_\mathrm{M}^0$, 
but generally, a larger region of attraction is achieved.

\section{Numerical Example}
The system model used in this example comes from \cite{lorenzen2019robust}. 
$A_n$ and $B_n$ are given as
\begin{align*} 
 	A_n=\begin{bmatrix}0.5&0.2\\-0.1&0.6\end{bmatrix}, B_n=\begin{bmatrix}0\\0.5\end{bmatrix}.
\end{align*} 

The unknown modeling error $\Delta P$ lies in 
\begin{align*} 
\mathbb{P}_\Delta = \left\{ \Delta P | \Delta P = \sum_{i=1}^3[\theta]_i \Delta P_i \right\}.
\end{align*}
with
\begin{align*} 
	&\Delta P_1 = \begin{bmatrix} 0.042&0&0 \\ 0.072&0.030&0 \end{bmatrix},  
	 \Delta P_2 = \begin{bmatrix}0.015&0.095&0 \\ 0.009&0.035&0 \end{bmatrix}, \\
	&\Delta P_3 = \begin{bmatrix}0&0&0.040 \\ 0&0&0.054\end{bmatrix}, 
	 \theta \in \Theta=\{ \theta \in \mathbb{R}^3 | \|\theta\|_\infty \leq 1 \}.
\end{align*}  
It should be noted that $\Delta P_i$ is not the vertex of $\mathbb{P}_\Delta$.

The actual value of $\Delta P$ is determined with $\theta^* = \begin{bmatrix}0.8 & 0.2 & -0.5\end{bmatrix}'$. 
Correspondingly, the multiplicative disturbance at time $t$ is  
\begin{align*} 
w_P(t)=\sum_{i=1}^3[\theta^*]_i \Delta P_i \begin{bmatrix}x(t)\\u(t)\end{bmatrix}.
\end{align*}

The set of admissible additional disturbance is 
\begin{align*}
	\mathbb{W}=\{w \in \mathbb{R}^2 |\|w\|_\infty\leq0.1\}.
\end{align*}

The set of admissible states and inputs is 
\begin{align*}
	\mathbb{Z}=\left\{\begin{bmatrix}x\\u\end{bmatrix}| x \in \mathbb{R}^2, u \in \mathbb{R}^1, \left | [x]_2 \right | \leq 30,\|u\|_\infty\leq1\right\}.
\end{align*}

Feedback gain is chosen as $K=\begin{bmatrix}0.0372  & -0.3261\end{bmatrix}$, which is determined by solving the algebraic Riccati equation with $A_n, B_n, Q=I,R=I$.
$\Psi$ is chosen as $I$.

$\mathbb{Z}_m^0$ is chosen to be an inner polytope of $\mathbb{B}^3$ with vertices  
\begin{align*}
	\begin{bmatrix}  sin(\theta_i)cos(\phi_j) & sin(\theta_i)sin(\phi_j) & cos(\theta_i) \end{bmatrix}' , i,j \in \mathbb{N}_{[1,5]}, 
\end{align*}
where $\theta_i$ and $\phi_j$ is chosen  uniformly between $0$ and $2\pi$.
With $\mathbb{Z}_m^0$, $\mathbb{P}_\Delta \mathbb{Z}_m^0$ is determined.

\begin{figure} 	
	\centering
	\includegraphics[width=8.8cm]{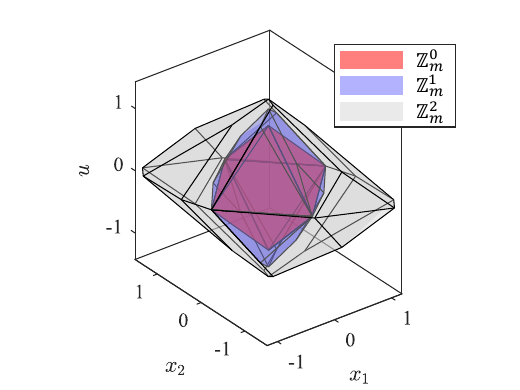}
	\caption{Compare of containers $\mathbb{Z}_m^0$, $\mathbb{Z}_m^1$, and $\mathbb{Z}_m^2$. }
	\label{fig:Z0_Z}
\end{figure}

\begin{figure} 	
	\centering
	\includegraphics[width=8.8cm]{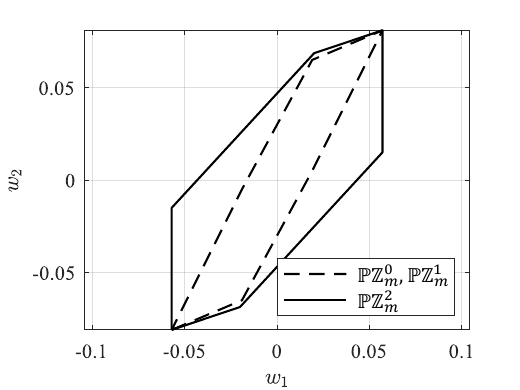}
	\caption{Compare of $\mathbb{PZ}_m^1$ and $\mathbb{PZ}_m^2$.}
	\label{fig:comparePZ}
\end{figure}

$\mathbb{Z}_m^1$ is determined by $\mathbb{Z}_m^1 = \{ z |\mathbb{P}_\Delta z \subseteq \mathbb{P}_\Delta \mathbb{Z}_m^0\}$.
$\mathbb{Z}_m^2$ is determined according to \ref{sec:optPZ0} with relaxed constraints on $\overline{\mathbb{W}}_\mathrm{M}^\dagger$.
$\mathbb{Z}_m^0$, $\mathbb{Z}_m^1$ and $\mathbb{Z}_m^2$ are shown in Fig. \ref{fig:Z0_Z} in red, blue, and gray, respectively.
The volume them are $1.3333,   1.9437$, and $4.2421$.
Both the volumes of  $\mathbb{Z}_m^1, \mathbb{Z}_m^2$ are greater than that of $\mathbb{Z}_m^0$.
Fig. \ref{fig:comparePZ} illustrates $\mathbb{P}_\Delta \mathbb{Z}_m^0$ and $\mathbb{P}_\Delta \mathbb{Z}_m^2$ with the areas with dashed and solid boundaries, respectively.  
The area of $\mathbb{P}_\Delta \mathbb{Z}_m^2$ is $0.0097$, over twice that of $\mathbb{P}_\Delta \mathbb{Z}_m^0$ which is $0.0046$.


\begin{figure} 	
 	\centering
 	\includegraphics[width=8.8cm]{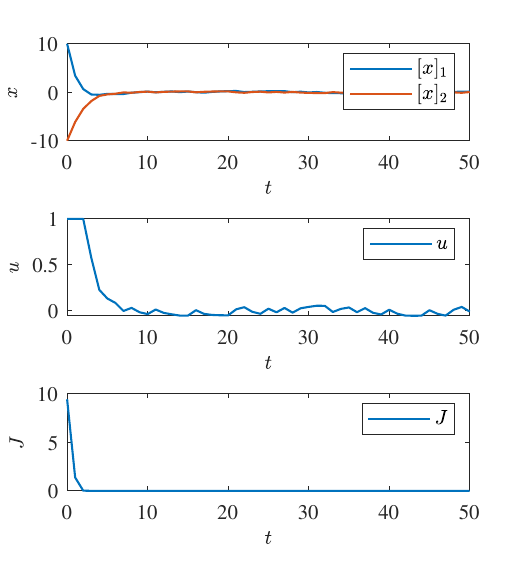}
 	\caption{Trajectory of states, input, and cost.}
 	\label{fig:trajectory}
 \end{figure}

 The optimization problems are implemented using YALMIP \cite{lofberg2004yalmip}.
 With prediction horizon $N=10$, $\mathbb{S}_\infty = \mathbb{S}_\infty(10)$, and initial condition $x(0)=[10;-10]$, 
 the trajectories of states, control input and cost are shown in Fig. \ref{fig:trajectory}.
 It can be seen that constraints on states and control inputs are satisfied and $x$ converge to a neighborhood of the origin.
 Meanwhile, the cost function is non-increasing during the convergence.
 
\begin{figure} 	
	\centering
	\includegraphics[width=8.8cm]{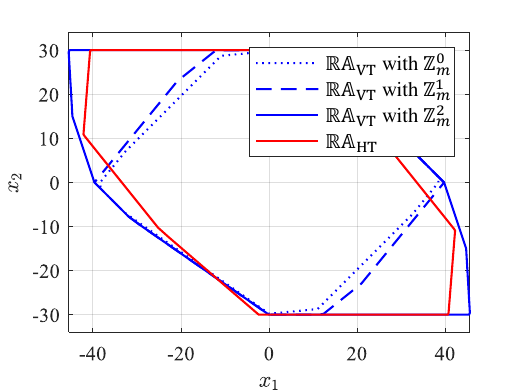}
	\caption{Compare of regions of attraction of different methods.}
	\label{fig:attractionRegion}
\end{figure}

Fig. \ref{fig:attractionRegion} illustrates the regions of attraction ($\mathbb{RA}$s) of the proposed CC-VT RMPC method and the HT-based RMPC method in \cite{lorenzen2019robust}. 
The regions of attraction of CC-VT RMPC method with $\mathbb{Z}_0, \mathbb{Z}_1$ and $\mathbb{Z}_2$ are notated as $\mathbb{RA}_{\mathrm{VT}}$ with $\mathbb{Z}_m^0, \mathbb{Z}_m^1$, and $\mathbb{Z}_m^2$, whose area are $2703.3, 2891.3$, and $3887.3$, respectively.
The method in \cite{lorenzen2019robust} utilizes homothetic tubes and its region of attraction is notated as $\mathbb{RA}_{\mathrm{HT}}$, whose area is $3554.6$.
It can be seen that with the optimized container, the region of attraction becomes larger. 
Meanwhile, the area of $\mathbb{RA}_{\mathrm{VT}}$ with $\mathbb{Z}_m^2$ is larger than that of $\mathbb{RA}_{\mathrm{HT}}$, showing that the proposed method is less conservative.

\begin{center}
	\begin{table*}[]
		\centering
		\caption{ compare of the online quadratic optimization problems in CC-VT RMPC and HT-based RMPC }
		\begin{tabular}{llll}
			\hline
			 & number of decision variables  & number of inequality constraints \\ \hline
			CC-VT RMPC                  & $N m + N$           & $N \big(N_{\mathbb{Z}_m}^{\mathrm{lcon}} + N_\mathbb{Z}^{\mathrm{lcon}} \big)+N_{\mathbb{S}_\infty}^{\mathrm{lcon}} + N$          \\
			CC-VT RMPC in this example  & 20                   & 308  \\
			HT-based RMPC                  & $N m +(N+1)(n+1)$   & 	
				$N N_{\mathbb{P}_\Delta}^{\mathrm{vert}}N_{\mathbb{X}_0}^{\mathrm{lcon}} + N N_\mathbb{Z}^{\mathrm{lcon}} + N_{\mathbb{S}_{2019}}^{\mathrm{lcon}} + N+1$ \\
			HT-based RMPC in this example  & 43                  & 724 \\ \hline
		\end{tabular}
		\label{tab:computation}
	\end{table*}
\end{center}

\section{Discussion}

\begin{figure} 	
	\centering
	\includegraphics[width=8.8cm]{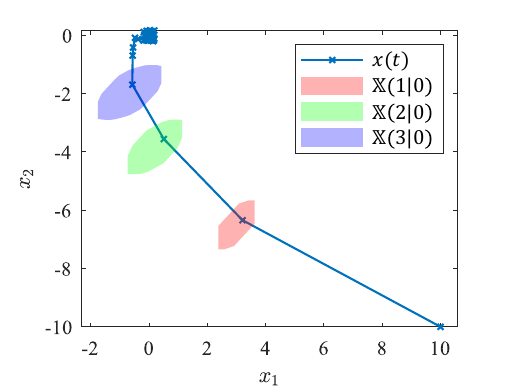}
	\caption{State trajectory and tubes $\mathbb{X}(k|0)$ with the proposed method, $k = 1,2,3.$ }
	\label{fig:VTtube}
\end{figure}

\begin{figure} 	
	\centering
	\includegraphics[width=8.8cm]{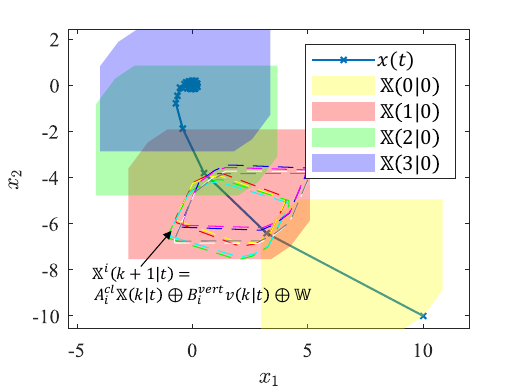}
	\caption{State trajectory and tubes $\mathbb{X}(k|0)$ with HT-based method \cite{lorenzen2019robust}, $k = 0,1,2,3.$}
	\label{fig:HTtube}
\end{figure}

The proposed CC-VT RMPC algorithm and HT-based RMPC algorithm take different methods to approximate admissible MD and connect adjacent tubes.
To illustrate the evolution of tubes, the trajectories of state and tubes under these two methods are shown in Fig. \ref{fig:VTtube} and Fig. \ref{fig:HTtube}, respectively.

CC-VT RMPC method approximates admissible MD explicitly.
In CC-VT RMPC method, tubes are expressed as 
\begin{align}
	\mathbb{X}(k|t) = \bar{x}(k|t) \oplus \overline{\mathbb{E}}(k|t).
\end{align}
By restricting  $\begin{bmatrix}\bar{x}(k|t) \\ \bar{u}(k|t) \end{bmatrix} \oplus \begin{bmatrix}I\\K\end{bmatrix} \overline{\mathbb{E}}(k|t)$ into  $\lambda(k|t) \mathbb{Z}_m$,
admissible MA is approximated explicitly and is supposed to be within $\lambda(k|t) \mathbb{P}_\Delta \mathbb{Z}_m$. 
Conservativeness is introduced because of the over-approximation.
The next tube $\bar{x}(k+1|t) \oplus \overline{\mathbb{E}}(k+1|t)$ is determined by 
\begin{align}
	\bar{x}(k+1|t) &= A_n \bar{x}(k|t) + B_n \bar{u}(k|t), \\
	\overline{\mathbb{E}}(k+1|t) &= A_n^{cl} \mathbb{X}(k|t) \oplus \mathbb{W} \oplus \lambda(k|t) \mathbb{P}_\Delta \mathbb{Z}_m.
\end{align}

HT-based RMPC approximates admissible MD implicitly. 
In HT-based RMPC method, tubes is expressed as $\mathbb{X}(k|t) =z(k|t) \oplus  \alpha(k|t) \mathbb{X}_0$ where $\mathbb{X}_0$ is the pre-defined shape of tubes.
A number of tubes $\mathbb{X}^i(k+1|t)$, illustrated by area with dashed boundary, can be determined by 
\begin{align}
	\mathbb{X}^i(k+1|t) = A^{cl}_i \mathbb{X}(k|t) \oplus B^\mathrm{vert}_i v(k|t) \oplus \mathbb{W},
\end{align}
where $ A^{cl}_i  = A_i^{\mathrm{vert}} +  B_i^{\mathrm{vert}} K$,  
$\begin{bmatrix} A_i^{\mathrm{vert}} &  B_i^{\mathrm{vert}}\end{bmatrix}$ is the vertex of $P_n \oplus \mathbb{P}_\Delta$,
$i \in \mathbb{N}_{ \left[1,N_{\mathbb{P}_\Delta}^{\mathrm{vert}} \right]}$.
At this step, MA is taken into consideration implicitly by traversing vertices of the set of admissible dynamics.
Then, $\mathbb{X}(k+1|t)$ are required to contain all those $\mathbb{X}^i(k+1|t)$, that is,
\begin{align} \label{eqn:HTconstraint}
	\mathbb{X}^i(k+1|t) \subseteq \bar{x}(k+1|t) \oplus  \alpha(k+1|t) \mathbb{X}_0.
\end{align}
The conservativeness is introduced here because $\mathbb{X}(k+1|t)$ is larger than $\mathbb{CH} \{\mathbb{X}^i(k+1|t)\}$. 

The numbers of decision variables and inequality constraints of the online optimization problems to be solved in CC-VT RMPC and HT-based RMPC are given in Table \ref{tab:computation}.
Computational complexity of a quadratic optimization problem is determined by the number of decision variables and the number of inequality constraints.
Since CC-VT RMPC involves a fewer number of decision variables and inequality constraints than HT-based RMPC does, 
the computational complexity of CC-VT RMPC is lower than that of HT-based RMPC. 

In CC-VT RMPC method, control variables $v(k|t)$ and the sizes of containers $\lambda(k|t)$ are decision variables, and $Nm+N$ decision variables are involved.
While in HT-based RMPC method, control variables $v(k|t)$,  the centers of tubes $z(k|t)$, and the sized of tubes $\alpha_i(k|t)$ are decision variables, 
thus, $N m +(N+1)(n+1)$ decision variables are involved.
VT-based RMPC involves $Nn+n+1$ more decision variables than CC-VT does.
In the above example, the numbers of decision variables of CC-VT RMPC and HT-based RMPC are 20 and 43, respectively.

In CC-VT RMPC method, additional constraints (\ref{eqn:addFeasibleConstraint}), admissibility constraints (\ref{eqn:feasibleConstraint}), terminal constraints (\ref{eqn:terminalConstraint}), and positiveness constraints (\ref{eqn:positivenessConstraint}) are imposed,
	 and $N \big(N_{\mathbb{Z}_m}^{\mathrm{lcon}} + N_\mathbb{Z}^{\mathrm{lcon}} \big)+N_{\mathbb{S}_\infty}^{\mathrm{lcon}} + N$ linear inequality constraints are involved.
In HT-based RMPC method, inclusion constraint are imposed to ensure (\ref{eqn:HTconstraint}). 
Along with admissibility constraints, terminal constraints, and positiveness constraints on $\alpha(k|t)$, 
the number of linear constraints in HT-based method amounts to $ N N_{\mathbb{P}_\Delta}^{\mathrm{vert}}N_{\mathbb{X}_0}^{\mathrm{lcon}} + N N_\mathbb{Z}^{\mathrm{lcon}} + N_{\mathbb{S}_{2019}}^{\mathrm{lcon}} + N+1$.
The first term $N N_{\mathbb{P}_\Delta}^{\mathrm{vert}}N_{\mathbb{X}_0}^{\mathrm{lcon}}$ may be very large, making VT-based RMPC contain more constraints than CC-VT RMPC does.
In the above example, $N_{\mathbb{P}_\Delta}^{\mathrm{vert}}=8$, and the numbers of inequality constraints of CC-VT RMPC and HT-based RMPC are 308 and 724, respectively.

When the dimension of system (\ref{eqn:actualSystem}) increases, CC-VT RMPC method gains greater advantages.
Firstly, HT-based method involves $Nn+n+1$ more variable, increasing linearly with the increase of system dimension.
Secondly, for systems in higher dimensions, the complexity of parameter uncertainty and that of the predefined shape of tube increase, thus, $ N_{\mathbb{P}_\Delta}^{\mathrm{vert}}$ and $N_{\mathbb{X}_0}^{\mathrm{lcon}}$ increase. 
This results in a dramatic increase of $N N_{\mathbb{P}_\Delta}^{\mathrm{vert}}N_{\mathbb{X}_0}^{\mathrm{lcon}}$.
Moreover, to obtain the terminal constraints in HT-based RMPC, a recursive algorithm is proposed where the projection from $2(n+1)$-dimension space to $(n+1)$-dimension space are required. When the dimension of system is large, the computational demand of the projection maybe be unacceptable.

\section{Conclusion}

In this paper, CC-VT RMPC method is proposed for the stabilization of constrained systems subject to both AD and MD.
AD resides in a known, bounded, and invariant set, but MD varies with state and control input, making it challenging to estimate and characterize MD.
This challenge is addressed by introducing concentric containers with free sizes and a fixed shape.
By restricting states and inputs into containers, admissible MDs are restricted in polytopes with varying sizes and a fixed shape.
Varying tubes are constructed then according to nominal dynamics and the knowledge of AD and MD, providing a tight boundary for reachable states.
Free sizes make the proposed method less conservative and the fixed shape makes it computationally efficient.
Moreover,  conservativeness is further reduced through the optimization of containers. 
Simulation results demonstrate the performance improvement achieved by the optimized containers, as well as the effectiveness of CC-VT RMPC method.
Compared to HT-based RMPC, CC-VT RMPC provides a novel solution to the stabilization problem of constrained systems subjected to both AD and MD, 
yielding a larger region of attraction while involving a fewer number of decision variables and constraints.
Future research interests include exploring a more flexible expression of containers and developing adaptive MPC with the VT-based structure.

\section{Appendix}
\setcounter{equation}{0}
\renewcommand\theequation{A.\arabic{equation}} 

\subsection{Proof of Theory 1}
\begin{proof}
	(i) This is ensured by (\ref{eqn:feasibleConstraint}) with $k=0$.
	
	(ii) The optimal solution to optimization problem (\ref{eqn:QP}) at $t$ is notated as 
	\begin{align} 
		{\bm{v}}^*({t}) &=[{v}^{*'}(0|t),{v}^{*'}(1|t),...,{v}^{*'}(N-1|t)]', \\
		\bm{\lambda}^*(t) &=[\lambda^*(0|t),\lambda^*(1|t),...,\lambda^*(N-1|t)]'. 
	\end{align}
	With $\bar{x}^*(0|t) = x(t)$,  $\bar{x}^*(k|t), k\in \mathbb{N}_{[1,N]}$ is determined according to (\ref{eqn:dynamicConstraint}) and (\ref{eqn:uBar})..
	With $\bm{\lambda}^*(t)$, $\overline{\mathbb{E}}(k|t), k \in \mathbb{R}_{[1,N]}$ is determined according to (\ref{eqn:E_E_W_WP}).
	
	At $t+1$, constraints (\ref{eqn:initialConstraints}) and (\ref{eqn:E_0}) are satisfied with $\bar{x}(0|t+1) = x(t+1)$ and $\overline{\mathbb{E}}(0|t+1) = \{ 0 \}$.
	With ${\bm{v}}({t+1})=[{v}^{*'}(1|t),{v}^{*'}(2|t),...,{v}^{*'}(N-1|t),0]'$ and
		 $\bm{\lambda}(t+1) = \bm{\lambda}(t+1) = [\lambda^*(1|t), \lambda^*(2|t), ..., \lambda^*(N-1|t), \lambda_\infty]$, 
	$\bar{\bm{x}}({t+1}) = [\bar{x}'(0|t+1),\bar{x}'(1|t+1),...,\bar{x}'(N|t+1)]'$ and $\overline{\mathbb{E}}(k|t+1)$ are determined accordingly.
	Here, $\lambda_\infty =\min_\gamma \{ \gamma |\mathbb{S}_\infty \subseteq \gamma\mathbb{X}_m \}$.
	
	Consider 
	\begin{align}
		\begin{aligned}
		\bar{x}(0|t+1) - \bar{x}^*(1|t) &= x(t+1) - \bar{x}^*(1|t) \\
										&\in \overline{\mathbb{E}}(1|t) 
										  = \lambda^*(0|t) \mathbb{P}_\Delta \mathbb{Z}_m \oplus \mathbb{W}.
		\end{aligned}
	\end{align}
	
	Meanwhile, with ${\bm{v}}({t+1})$ and $k\in \mathbb{R}_{[0,N-1]}$, we have 
	$\bar{x}(k+1|t+1) - \bar{x}^*(k+2|t) = A_n^{cl}(\bar{x}(k|t+1) - \bar{x}^*(k+1|t))$.
	Thus, $\bar{x}(k|t+1) - \bar{x}^*(k+1|t) \in (A_n^{cl})^k \overline{\mathbb{E}}(1|t) $.
	
	Further, 
	\begin{align}
		\begin{aligned}
		&\overline{\mathbb{E}}(k+1|t) \ominus (A_n^{cl})^k \overline{\mathbb{E}}(1|t) \\
		&~~~~= \sum_{i=0}^{k}\lambda^* (k-i|t)(A_n^{cl})^i \mathbb{P}_\Delta \mathbb{Z}_m  \oplus \sum_{i=0}^{k}(A_n^{cl})^i\mathbb{W} \\
		&~~~~~~~~  \ominus (A_n^{cl})^k (\lambda^*(0|t) \mathbb{P}_\Delta \mathbb{Z}_m \oplus \mathbb{W} ) \\
		&~~~~= \sum_{i=0}^{k-1}\lambda^* (k-i|t)(A_n^{cl})^i \mathbb{P}_\Delta \mathbb{Z}_m  \oplus \sum_{i=0}^{k-1}(A_n^{cl})^i\mathbb{W} \\
		&~~~~= \sum_{i=0}^{k-1}\lambda (k-1-i|t+1)(A_n^{cl})^i \mathbb{P}_\Delta \mathbb{Z}_m  \oplus \sum_{i=0}^{k-1}(A_n^{cl})^i\mathbb{W} \\
		&~~~~= \overline{\mathbb{E}}(k|t+1).
		\end{aligned}
	\end{align}
	Meanwhile, $\bar{x}^*(N|t) \in \mathbb{S}_\infty \ominus \mathbb{E}(N|t) \subseteq \mathbb{X}_{xu} \ominus \mathbb{E}(N|t)$ implies 
	\begin{align}
		\begin{bmatrix}\bar{x}^*(N|t)\\K\bar{x}^*(N|t) \end{bmatrix} \in \mathbb{Z} \ominus\begin{bmatrix}I\\K\end{bmatrix}\overline{\mathbb{E}}(N|t).
	\end{align}
	
	Then, together with (\ref{eqn:feasibleConstraint}), we have 
	\begin{align}
	\begin{aligned}
		&\begin{bmatrix}\bar{x}(k|t+1)\\K\bar{x}(k|t+1)+v(k|t+1)\end{bmatrix} \\
		&\subseteq
		\begin{bmatrix}\bar{x}^*(k+1|t)\\K\bar{x}^*(k+1|t)+v^*(k+1|t)\end{bmatrix} \oplus \begin{bmatrix}I\\K\end{bmatrix} (A_n^{cl})^k \overline{\mathbb{E}}(1|t) \\
		&\subseteq
		\mathbb{Z} \ominus\begin{bmatrix}I\\K\end{bmatrix}\overline{\mathbb{E}}(k+1|t) \oplus \begin{bmatrix}I\\K\end{bmatrix}(A_n^{cl})^k \overline{\mathbb{E}}(1|t) \\
		&\subseteq \mathbb{Z} \ominus\begin{bmatrix}I\\K\end{bmatrix} \overline{\mathbb{E}}(k|t+1).
	\end{aligned}
	\end{align}
	
	This means, at $t+1$, constraint (\ref{eqn:feasibleConstraint}) are satisfied.
	
	$\bar{x}^*(N|t) \in \mathbb{S}_\infty \ominus \mathbb{E}(N|t) \subseteq \gamma_\infty \mathbb{X}_{m} \ominus \mathbb{E}(N|t)$ implies 
	\begin{align}
		\begin{bmatrix}\bar{x}^*(N|t)\\K\bar{x}^*(N|t) \end{bmatrix} \in \gamma_\infty \mathbb{Z}_{m} \ominus \begin{bmatrix}I\\K\end{bmatrix} \overline{\mathbb{E}}(N|t).
	\end{align}
	
	Similarly, together with (\ref{eqn:addFeasibleConstraint}), we have 
	\begin{align}
		\begin{aligned}
			&\begin{bmatrix}\bar{x}(k|t+1)\\K\bar{x}(k|t+1)+v(k|t+1)\end{bmatrix} \\
			&\subseteq
			\begin{bmatrix}\bar{x}^*(k+1|t)\\K\bar{x}^*(k+1|t)+v^*(k+1|t)\end{bmatrix} \oplus \begin{bmatrix}I\\K\end{bmatrix} (A_n^{cl})^k \overline{\mathbb{E}}(1|t) \\
			&\subseteq
			\lambda^*(k+1|t) \mathbb{Z}_m \ominus\begin{bmatrix}I\\K\end{bmatrix}\overline{\mathbb{E}}(k+1|t) \oplus \begin{bmatrix}I\\K\end{bmatrix}(A_n^{cl})^k \overline{\mathbb{E}}(1|t) \\
			&\subseteq \lambda^*(k+1|t) \mathbb{Z}_m \ominus\begin{bmatrix}I\\K\end{bmatrix} \overline{\mathbb{E}}(k|t+1).
		\end{aligned}
	\end{align}
	
		This means, at $t+1$, constraint (\ref{eqn:addFeasibleConstraint}) are satisfied.
	
	At the same time, 
	\begin{align}
		\begin{aligned}
		&\bar{x}(N|t+1) = A_n^{cl} \bar{x}(N-1|t+1) \\
		&~~~~\in A_n^{cl} ( \bar{x}^*(N|t) \oplus (A_n^{cl})^{N-1} \mathbb{E}_0(t) ) \\
		&~~~~\subseteq A_n^{cl} ( \mathbb{S}_\infty \ominus \overline{\mathbb{E}}(N|t) \oplus (A_n^{cl})^{N-1} \overline{\mathbb{E}}(1|t) ) \\
		&~~~~\subseteq A_n^{cl} ( \mathbb{S}_\infty \ominus \overline{\mathbb{E}}(N-1|t+1) ) \\
		&~~~~\subseteq \mathbb{S}_\infty \ominus \mathbb{W} \ominus \gamma_\infty\overline{\mathbb{W}}_p \ominus A_n^{cl} \overline{\mathbb{E}}(N-1|t+1)) \\
		&~~~~ \subseteq \mathbb{S}_\infty \ominus \overline{\mathbb{E}}(N|t+1).
		\end{aligned}
	\end{align}
	
	This means, at $t+1$, constraint (\ref{eqn:terminalConstraint}) are satisfied.
	
	Thus, the solution with ${\bm{v}}({t+1})$ and $\bm{\lambda}(t+1)$ defined above is feasible  to the optimization problem (\ref{eqn:QP}) and (ii) holds.
	
	(iii) With the feasible solution given in (ii), the corresponding cost at $t+1$ is 
	\begin{align}
		J(\bm{v}(t+1), \bm{\lambda}(t+1)) = \sum_{k=1}^{N-1}\|v^*(k|t)\|_\psi^2.
	\end{align}
	Then,
	\begin{align}
		\begin{aligned}
			&~~~~	  J(\bm{v}^*(t+1), \bm{\lambda}^*(t+1)) - J(\bm{v}^*(t), \bm{\lambda}^*(t)) \\
		    &\leq J(\bm{v}(t+1), \bm{\lambda}(t+1))     - J(\bm{v}^*(t), \bm{\lambda}^*(t))  \\
			&= -\|v^*(0|t)\|_\psi^2.
		\end{aligned}
	\end{align}
	Since $\psi$ is positive definite,  $J(\bm{v}^*(t), \bm{\lambda}^*(t))$ is monotone non-increasing and non-negative. 
	Thus, $J(\bm{v}^*(t), \bm{\lambda}^*(t))$ converges to a constant, $v^*(0|t)$ converges to $0$, and $u(t)$ converges to $Kx(t)$.
	This means 
	\begin{align}
		\begin{bmatrix} x(\infty)\\K x(\infty) + v^*(0|\infty) \end{bmatrix} = \begin{bmatrix} I\\K   \end{bmatrix} x(\infty).
	\end{align}
	According to (i), 
	$\begin{bmatrix}I\\K\end{bmatrix} x(\infty) \in\mathbb{Z}$, thus, $x(\infty) \in \mathbb{X}_{xu}$.
	$\mathbb{X}_{xu}$ contains the origin and (iii) holds.
\end{proof}

\subsection{Proof of Theory 2}
\begin{proof}
	$\forall z \in \mathbb{Z}_m^1$, $\mathbb{P}_\Delta z \subseteq  \mathbb{P}_\Delta \mathbb{Z}_m^1$, 
	Thus, $\mathbb{Z}_m^1 \subseteq \mathbb{Z}_m^2$.
	
	With  $\mathbb{Z}_m^1 \subset \mathbb{Z}_m^2$, we have $\mathbb{X}_m^1 \subset \mathbb{X}_m^2$, where
	\begin{align}
		\mathbb{X}_{m}^j = \Big\{x|\begin{bmatrix} x \\ Kx \end{bmatrix} \in \mathbb{Z}_m^j\Big\}, j=1,2. 
	\end{align}

	With $\mathbb{Z}_m = \mathbb{Z}_m^1$ and $\mathbb{Z}_m = \mathbb{Z}_m^2$, 
	the corresponding $\overline{\mathbb{E}}(0|t)$ are $\overline{\mathbb{E}}_1(0|t)=\overline{\mathbb{E}}_2(k|t)=\{ 0\}$.
	The corresponding $\overline{\mathbb{E}}(k|t), k\in\mathbb{Z}_{[1,N]},$ are notated as $\overline{\mathbb{E}}_1(k|t)$ and $\overline{\mathbb{E}}_2(k|t)$, receptively and are expressed as
	\begin{align}  
		\begin{aligned}
			 \overline{\mathbb{E}}_j(k|t) &= \sum_{i=0}^{k-1}(A_n^{cl})^i\mathbb{W} \\
			& \oplus\sum_{i=0}^{k-1}\lambda(k-1-i|t)(A_n^{cl})^i \mathbb{P}_\Delta \mathbb{Z}_m^j, j=1,2.
		\end{aligned}
	\end{align}
	
	Since $\mathbb{P}_\Delta \mathbb{Z}_m^2 = \mathbb{P}_\Delta \mathbb{Z}_m^1$, we have $\overline{\mathbb{E}}_2(k|t) = \overline{\mathbb{E}}_1(k|t)$.	
	
	Further,
	\begin{align}
		\begin{aligned}
			\lambda(k|t) \mathbb{Z}_m^1 \ominus \Pi \overline{\mathbb{E}}_1(k|t) &= \lambda(k|t) \mathbb{Z}_m^1 \ominus \Pi \overline{\mathbb{E}}_2(k|t) \\
			&\subset \lambda(k|t) \mathbb{Z}_m^2 \ominus \Pi \overline{\mathbb{E}}_2(k|t), \\
			\mathbb{Z} \ominus \Pi \overline{\mathbb{E}}_1(k|t) 			&= \mathbb{Z} \ominus \Pi \overline{\mathbb{E}}_2(k|t),
		\end{aligned}
	\end{align}
	where 
	\begin{align}
		\Pi = \begin{bmatrix}I\\K\end{bmatrix}.
	\end{align}
	
	With $\mathbb{Z}_m = \mathbb{Z}_m^1$ and $\mathbb{Z}_m = \mathbb{Z}_m^2$, 
	the corresponding $\mathbb{S}_n$ are notated as $\mathbb{S}_n^1$ and $\mathbb{S}_n^2$, respectively.
	Since $\mathbb{X}_m^1 \subset \mathbb{X}_m^2$, we have 
		$\mathbb{S}_0^1 = \gamma_\infty \mathbb{X}_m^1 \cap \mathbb{X}_{xu} \subseteq \gamma_\infty \mathbb{X}_m^2 \cap \mathbb{X}_{xu} = \mathbb{S}_0^2$
	
	Suppose $\mathbb{S}_{n-1}^1 \subseteq \mathbb{S}_{n-1}^2$, then $\forall x \in \mathbb{S}_n^1$, we have $x \in \mathbb{S}_0^1 \subseteq {\mathbb{S}}_{0}^2 $ and 
	\begin{align}
		\begin{aligned}
			A_n^{cl} x 
			&\in       \mathbb{S}_{n-1}^1 \ominus \mathbb{W} \ominus \gamma_\infty \mathbb{P}_\Delta \mathbb{Z}_m^1 \\
			&= 	       \mathbb{S}_{n-1}^1 \ominus \mathbb{W} \ominus \gamma_\infty \mathbb{P}_\Delta \mathbb{Z}_m^2 \\
			&\subseteq \mathbb{S}_{n-1}^2 \ominus \mathbb{W} \ominus \gamma_\infty \mathbb{P}_\Delta \mathbb{Z}_m^2.
		\end{aligned}
	\end{align}
	Thus, $\mathbb{S}_{n}^1 \subseteq \mathbb{S}_{n}^2$.
	Then, with 	$\mathbb{S}_0^1(\gamma_0) \subseteq \mathbb{S}_0^1(\gamma_0)$, we have $\mathbb{S}_{n}^1 \subseteq \mathbb{S}_{n}^2$.
	This implies 
	\begin{align}
	   	  \mathbb{S}_{\infty}^1 \ominus \overline{\mathbb{E}}_1(k|t) 
		= \mathbb{S}_{\infty}^1 \ominus \overline{\mathbb{E}}_2(k|t)
\subseteq \mathbb{S}_{\infty}^2 \ominus \overline{\mathbb{E}}_2(k|t).
	\end{align}
	
	Thus, all feasible solutions to optimization problem (\ref{eqn:QP}) with $\mathbb{Z}_m = \mathbb{Z}_m^1$ is also feasible to (\ref{eqn:QP}) with $\mathbb{Z}_m = \mathbb{Z}_m^2$.
	
	On the contrary, since $\mathbb{Z}_m^1 \subset \mathbb{Z}_m^2$, there exists solutions such that constraint (\ref{eqn:addFeasibleConstraint}) is satisfied with $\mathbb{Z}_m = \mathbb{Z}_m^2$ but is not satisfied with $\mathbb{Z}_m = \mathbb{Z}_m^1$.
	
	Thus, if $\mathbb{Z}_m^1 \subset \mathbb{Z}_m^2$, $\mathbb{Z}_m^2$ is better than $\mathbb{Z}_m^1$.	
\end{proof}

\section*{References}
\bibliographystyle{IEEEtran}
\bibliography{biblio_RMPC_TAC}

\end{document}